\documentclass[a4paper,number,5p,twocolumn]{elsarticle}

\usepackage[english]{babel}
\usepackage[utf8]{inputenc}
\usepackage{graphicx}
\usepackage[usenames,dvipsnames]{xcolor}
\usepackage{bm}
\usepackage{array}
\usepackage{amsfonts}
\usepackage{amssymb}
\usepackage{amsmath}
\usepackage{epstopdf}
\usepackage{soul}
\usepackage[normalem]{ulem}
\usepackage[pagewise,switch,mathlines]{lineno}
\usepackage{setspace}
\usepackage{csquotes}
\usepackage{algorithm}
\usepackage{algpseudocode}
\usepackage[
    unicode,
    breaklinks=true,
    hypertexnames=false,
    colorlinks=false,
    citecolor=black,
    filecolor=black,
    linkcolor=black,
    urlcolor=black
]{hyperref}
\usepackage{natbib}
\usepackage{mathtools}
\usepackage{interval}
\usepackage{multirow}
\usepackage{paralist}
\usepackage{xifthen}

\newcommand{\Fref}[1]{Fig.~\ref{#1}}
	
\newcommand{\Eref}[1]{Eq.~(\ref{#1})}

\newcommand{\Sref}[1]{{Section~\ref{#1}}}

\newcommand{\tile}[1]{\ensuremath{\mathcal{T}\ifthenelse{\isempty{#1}}{}{_{#1}} }}
\newcommand{\tileset}{\ensuremath{\mathcal{S}}}

\newcommand{\particle}{\ensuremath{\mathcal{P}}}
\newcommand{\particleBoundary}{\ensuremath{\Gamma^\particle}}
\newcommand{\radius}{\ensuremath{r}}
\newcommand{\admissibleDomain}{\ensuremath{\mathcal{A}}}

\newcommand{\LS}{\ensuremath{\mathcal{L}}}
\newcommand{\LSp}{\ensuremath{\LS^{\particle}}}
\newcommand{\LSd}{\ensuremath{\LS^{\domain}}}
\newcommand{\LSsSimple}{\ensuremath{\LS^{\tileset}}}
\newcommand{\LSs}[1]{\ensuremath{\LSsSimple\ifthenelse{\isempty{#1}}{}{_{\text{\romannumber{#1}}}}}}
\newcommand{\LSdi}[1]{\ensuremath{\LSd_{#1}}}

\newcommand{\volfrac}{\ensuremath{\phi}}

\newcommand{\Stwo}{\ensuremath{S_2}}
\newcommand{\StwoSec}{\ensuremath{\hat{S}_2}}

\newcommand{\romannumber}[1]{\uppercase\expandafter{\romannumeral #1\relax}}

\mathchardef\mhyphen="2D

\newcommand{\atx}{\ensuremath{(\tens{x})}}

\newcommand{\domain}{\Omega}

\newcommand{\norm}[1]{\left\lVert#1\right\rVert}

\newcommand{\tens}[1]{\boldsymbol{#1}}					

\newcommand{\x}{\tens{x}} 								

\newcommand{\y}{\ensuremath{\tens{y}}}

\DeclareGraphicsExtensions{.pdf,.eps,.png,.jpg}	
\graphicspath{ {./} }

\soulregister\cite7
\soulregister\ref{7}
\soulregister\pageref7

\makeatletter
\providecommand{\doi}[1]{%
	\begingroup
	\let\bibinfo\@secondoftwo
	\urlstyle{rm}%
	\href{http://dx.doi.org/#1}{%
		doi:\discretionary{}{}{}%
		\nolinkurl{#1}%
	}%
	\endgroup
}
\makeatother

\newcommand*\patchAmsMathEnvironmentForLineno[1]{%
	\expandafter\let\csname old#1\expandafter\endcsname\csname #1\endcsname
	\expandafter\let\csname oldend#1\expandafter\endcsname\csname end#1\endcsname
	\renewenvironment{#1}%
	{\linenomath\csname old#1\endcsname}%
	{\csname oldend#1\endcsname\endlinenomath}}%
\newcommand*\patchBothAmsMathEnvironmentsForLineno[1]{%
	\patchAmsMathEnvironmentForLineno{#1}%
	\patchAmsMathEnvironmentForLineno{#1*}}%
\AtBeginDocument{%
	\patchBothAmsMathEnvironmentsForLineno{equation}%
	\patchBothAmsMathEnvironmentsForLineno{align}%
	\patchBothAmsMathEnvironmentsForLineno{flalign}%
	\patchBothAmsMathEnvironmentsForLineno{alignat}%
	\patchBothAmsMathEnvironmentsForLineno{gather}%
	\patchBothAmsMathEnvironmentsForLineno{multline}%
}

\makeatletter
\def\ps@pprintTitle{
	\let\@oddhead\@empty
	\let\@evenhead\@empty
	\def\@oddfoot{{\small© 2020. This manuscript version is made available under the \href{http://creativecommons.org/licenses/by-nc-nd/4.0/}{ CC-BY-NC-ND 4.0 license}.}\hfill{}}%
	\let\@evenfoot\@oddfoot
}
\makeatother

\begin{document}


\begin{frontmatter}

\title{Level-set based design of Wang tiles for modelling complex microstructures\tnoteref{t1}}
\tnotetext[t1]{Author's post-print version of the article manuscript published in \mbox{\textit{Computer-Aided Design}}\\\href{https://doi.org/10.1016/j.cad.2020.102827}{DOI: 10.1016/j.cad.2020.102827}.}

\author[mech,exp]{Martin Do\v{s}k\'{a}\v{r}}
\ead{martin.doskar@fsv.cvut.cz}
\author[mech,utia]{Jan Zeman}
\ead{jan.zeman@cvut.cz}
\author[mech]{Daniel Rypl}
\ead{daniel.rypl@fsv.cvut.cz}
\author[exp]{Jan Nov\'{a}k}
\ead{novakja@fsv.cvut.cz}
\address[mech]{Department of Mechanics, Faculty of Civil Engineering, Czech Technical University in Prague, Th\'{a}kurova 7, \mbox{166 29 Prague 6}, Czech Republic}
\address[utia]{The Institute of Information Theory and Automation, Academy of Sciences of the Czech Republic, Pod Vodárenskou věží 4, \mbox{182 08 Prague 8}, Czech Republic}
\address[exp]{Experimental centre, Faculty of Civil Engineering, Czech Technical University in Prague, Th\'{a}kurova 7, \mbox{166 29 Prague 6}, Czech Republic}

\journal{arXiv.org}

\begin{abstract}
	
	Microstructural geometry plays a critical role in the response of heterogeneous materials. Consequently, methods for generating microstructural samples are increasingly crucial to advanced numerical analyses. 
	We extend Sonon et al.'s unified framework, developed originally for generating particulate and foam-like microstructural geometries of Periodic Unit Cells, to non-periodic microstructural representations based on the formalism of Wang tiles. 
	This formalism has been recently proposed in order to generalize the Periodic Unit Cell approach, enabling a fast synthesis of arbitrarily large, stochastic microstructural samples from a handful of domains with predefined microstructural compatibility constraints. However, a robust procedure capable of designing complex, three-dimensional, foam-like and cellular morphologies of Wang tiles has not yet been proposed. This contribution fills the gap by significantly broadening the applicability of the tiling concept.
	
	Since the original Sonon et al.'s framework builds on a random sequential addition of particles enhanced with an implicit representation of particle boundaries by the level-set field, we first devise an analysis based on a connectivity graph of a tile set, resolving the question where a particle should be copied when it intersects a tile boundary. Next, we introduce several modifications to the original algorithm that are necessary to ensure microstructural compatibility in the generalized periodicity setting of Wang tiles.
	Having established a universal procedure for generating tile morphologies, we compare strictly aperiodic and stochastic sets with the same cardinality in terms of reducing the artificial periodicity in reconstructed microstructural samples. We demonstrate the superiority of the vertex-defined tile sets for two-dimensional problems and illustrate the capabilities of the algorithm with two- and three-dimensional examples.
	
\end{abstract}

\begin{keyword}
	Microstructure generation, Random heterogeneous materials, Wang tiling, Level sets, Foam microstructures, Cellular microstructures
\end{keyword}

\end{frontmatter}



\section{Introduction}
\label{sec:introduction}

The geometrical details of a material composition drive many macroscopic phenomena such as crack initiation and propagation~\citep{nguyen_homogenization-based_2011} or meta-behaviour~\citep{barchiesi_mechanical_2018}. Advanced computational strategies thus tend to incorporate knowledge of material microstructures\footnote{We use the term \enquote{microstructure} in a broader sense, without referring to a specific scale length.}	in order to enhance their predictive power.~\cite{matous_review_2017}

When micro and macro scales are well separated, a microstructural response is typically up-scaled through homogenization methods. Albeit computationally intensive,
numerical homogenization methods~\citep{matous_review_2017} now supersede analytical methods~\citep{torquato_random_2002} because of their ability to handle complex microstructural geometries and non-linear problems. With increasing computational power, numerical models that fully resolve the microstructural geometry in the whole macroscopic domain, e.g.~\citep{nguyen_large-scale_2017,lloberas-valls_domain_2011}, or in regions of interest~\citep{akbari_rahimabadi_scale_2015,lloberas-valls_multiscale_2012}, have emerged, addressing problems without clear scale separation.

For both strategies, however, accuracy critically depends on the representativeness of the provided microstructural geometry, accenting the crucial role of microstructure modelling in multi-scale approaches.

Compared to materials with regular microstructures, characterized entirely by Periodic Unit Cells (PUCs), modelling random heterogeneous materials is more intricate; any finite-size representation automatically implies information loss. The optimal microstructure representation should capture intrinsic randomness and fluctuations in a microstructure while remaining computationally tractable (for numerical homogenization) or inexpensive to construct (for fully resolved simulations).
\begin{figure}
	\centering
	\includegraphics[width=0.85\columnwidth]{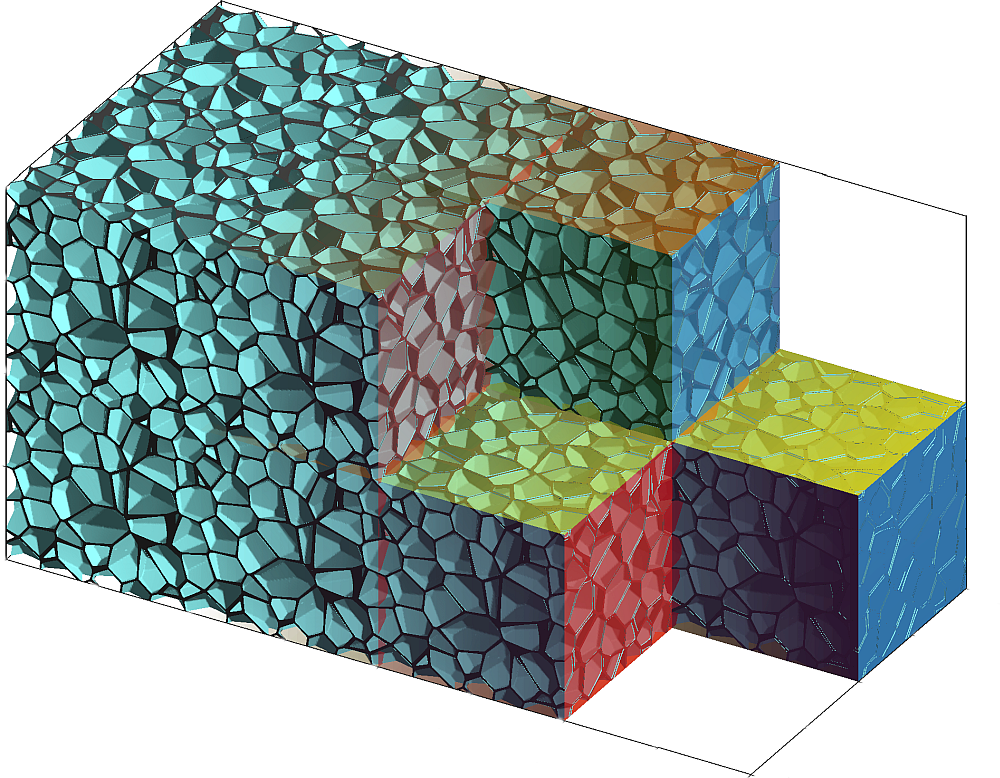}
	\caption{Illustration of a synthesized sample of a closed-foam microstructure obtained as a $4\times2\times2$ tiling assembled from a set of pre-generated Wang cubes. Individual face codes that encode the geometrical continuity of individual tiles and play the role of compatibility constraints during a tiling assembly are shown in semi-transparent colours.}
	\label{fig:advertisement}
\end{figure}

\subsection{State-of-the-art in modelling random microstructures}
\label{sec:state_of_the_art}

One of the widely adopted approaches for modelling heterogeneous materials rests on an extension of PUC generated such that its spatial statistics match that of a reference microstructure. 
This procedure appears in the literature under various names such as Statistically Optimal Representative Unit Cell~\citep{lee_three-dimensional_2009}, Repeating Unit Cell~\citep{yang_new_2018}, Statistically Similar Representative Volume Element~\citep{balzani_construction_2014}, or Statistically Equivalent Periodic Unit Cell (SEPUC)~\citep{zeman_random_2007}, among others.
The spatial statistics involved range from Minkowski functionals~\citep{scheunemann_design_2015} to multi-point probability functions~\citep{torquato_random_2002}, out of which the two-point probability~\citep{jiao_modeling_2007,zeman_numerical_2001,rozman_efficient_2001}, two-point cluster~\citep{jiao_superior_2009}, and lineal path~\citep{lu_lineal-path_1992,kumar_using_2006,zeman_random_2007,havelka_compression_2016} functions are the most frequently used.

Following Povirk's seminal work~\citep{povirk_incorporation_1995}, the majority of cell representations are generated using optimization procedures, minimizing the discrepancy between the statistical characterization of the reference microstructure and its compressed, PUC-like representation. The particular choice of optimization algorithm currently varies with several options including simulated annealing~\citep{yeong_reconstructing_1998,kumar_using_2006,zeman_random_2007}, genetic~\citep{basanta_using_2005,kumar_reconstruction_2008,lee_three-dimensional_2009,yang_new_2018} and gradient~\citep{povirk_incorporation_1995,fullwood_gradient-based_2008}, or phase-recovery~\citep{fullwood_microstructure_2008} algorithms.

The second approach to microstructure generation utilizes reference samples of the microstructure. New realizations are then obtained with a Markovian process, taking individual voxels~\citep{liu_random_2015} or a patch of voxels~\citep{tahmasebi_cross-correlation_2013} from the provided reference samples according to the proximity of the spatial statistics computed for their surroundings. Alternatively, searching for statistics proximity can be replaced with a classification tree-based supervised learning model~\citep{bostanabad_stochastic_2016}.

The previous two approaches suffer from high computational costs related either to optimization or to training the learning models. The applicability of their outputs is also sensitive to the spatial statistics considered, attesting to the ill-conditioning of the microstructure reconstruction problem itself. Achieving a good match in selected statistics does not automatically guarantee similar overall behaviour; for instance, \citet{biswal_quantitative_1999} demonstrated that realizations with similar two-point probability functions could have significantly different percolation characteristics that govern overall transport properties.

Complementary to the statistics-informed methods, a third approach to generating microstructural realizations relies on meta-modelling the genesis of a microstructure. These methods range in complexity and include the Monte-Carlo Potts~\citep{saito_monte_1992} and phase field models~\citep{krill_iii_computer_2002} of grain growth; sedimentation-and-compaction models~\citep{biswal_quantitative_1999}; and various particle packing algorithms, e.g.~\citep[and references therein]{falco_generation_2017,alsayednoor_evaluating_2016}, based on either Random Sequential Adsorption (RSA)\footnote{{In its simplest setting of packing circular or spherical particles, RSA corresponds to the Dart Throwing Algorithm~\cite{mitchell_generating_1987}.}}~\citep{cooper_random-sequential-packing_1988,segurado_numerical_2002} or molecular dynamics~\citep{lubachevsky_geometric_1990,donev_neighbor_2005,ghossein_random_2013}.

The relevance of packing algorithms extends beyond simple particle-matrix microstructures because the resulting packings often serve as initial seeds for tessellation-based models applicable to polycrystals~\citep{de_giorgi_aluminium_2010,chen_effects_2015}, foams~\citep{alsayednoor_evaluating_2016}, and cell tissues~\citep{mebatsion_microscale_2006,chakraborty_adaptive_2013}. Due to its straightforward implementation, Vorono\"{i} tessellation is the most common choice; however, the resulting geometry is oversimplified for many materials. For instance, Vorono\"{i}-based models overestimate overall stiffness for high porosity foams~\citep{doskar_jigsaw_2016}. The curvature of the cell walls~\citep{de_giorgi_aluminium_2010,simone_effects_1998} and heterogeneity in cell size~\citep{alsayednoor_evaluating_2016,chen_effects_2015} and wall thickness~\citep{chen_effects_2015} must be additionally introduced to obtain realistic geometries. Similar effects can be achieved by modifying the distance measure used during tessellation, e.g. models based on the Laguerre variant generate microstructures with multi-mode cell size distribution~\citep{lyckegaard_use_2011,chen_effects_2015,alsayednoor_evaluating_2016,falco_generation_2017}. Inspired by Laguerre tessellation, \citet{chakraborty_adaptive_2013} proposed Adaptive Quadratic Vorono\"{i} Tessellation, attributing a distinct anisotropic metric to each seed and thus allowing for additional control over the resulting geometry.

The original RSA method~\cite{mitchell_generating_1987} suffers from $\mathcal{O}\left(N^2\right)$ complexity for $N$ particles due to overlap checks and is impractical for generating large, densely packed systems. Consequently, several accelerations have been proposed.
For Dart Throwing Algorithm~{\citep{mitchell_generating_1987}}, which is a simplified case of RSA with equisized circular/spherical particles, {\citet{dunbar_spatial_2006}} introduced a scalloped sector representation of non-overlap guaranteed regions. In the same year, {\citet{jones_efficient_2006}} proposed an alternative bookkeeping of the regions based on an adaptively updated Vorono\"{i} tessellation. Moreover, both approaches utilize a tree data structure and improve the algorithm complexity to $\mathcal{O}(N\log N)$. For general RSA, {\citet{yang_new_2018}} proposed an acceleration based on a combination of a spline description of particle shapes and hierarchically refined bounding boxes of each particle.
Recently, \citet{sonon_unified_2012} introduced a method building on an implicit, level-set based description of particle shapes, achieving $\mathcal{O}(N)$ complexity (see Section 3.3 in~\citep{sonon_unified_2012}). 
Moreover, Sonon et al.'s method readily facilitates generating complex microstructures using linear combinations of the nearest neighbour distance functions and dedicated morphing operations~\citep{sonon_unified_2012,sonon_level-set_2013,sonon_advanced_2015,massart_level_2017}. In a sense, this approach introduces the anisotropic pseudo-metrics\footnote{Level-set description based on the signed distance to the $n$th nearest particle boundary does not fulfil all metric criteria.} of~\citep{chakraborty_adaptive_2013} in a geometrically-motivated way by considering arbitrarily-shaped particles. As a result, Sonon et al.'s method enables refined control over generated microstructure unattainable with standard Vorono\"{i} or Laguerre tessellations; see \Sref{sec:morphing}.

Albeit significantly faster than RSA or optimization-based approaches, the latter method still starts anew every time an additional realization is required, imposing overhead on, e.g., investigations of the Representative Volume Element (RVE) size that require multiple microstructural samples to be generated, see~\citep{kanit_determination_2003,gitman_representative_2007,dirrenberger_towards_2014}. Alternatively, larger microstructural realizations can be assembled from (SE)PUC; however, such construction introduces non-physical, long-range, periodic artefacts in a microstructural geometry and its local response.

\subsection{Wang tiling in microstructure modelling}
\begin{figure*}[ht!]
	\centering
	\setlength{\tabcolsep}{4pt}
	\begin{tabular}{cp{0.28\textwidth}cp{0.28\textwidth}cp{0.28\textwidth}}
		(a) & \includegraphics[width=0.25\textwidth]{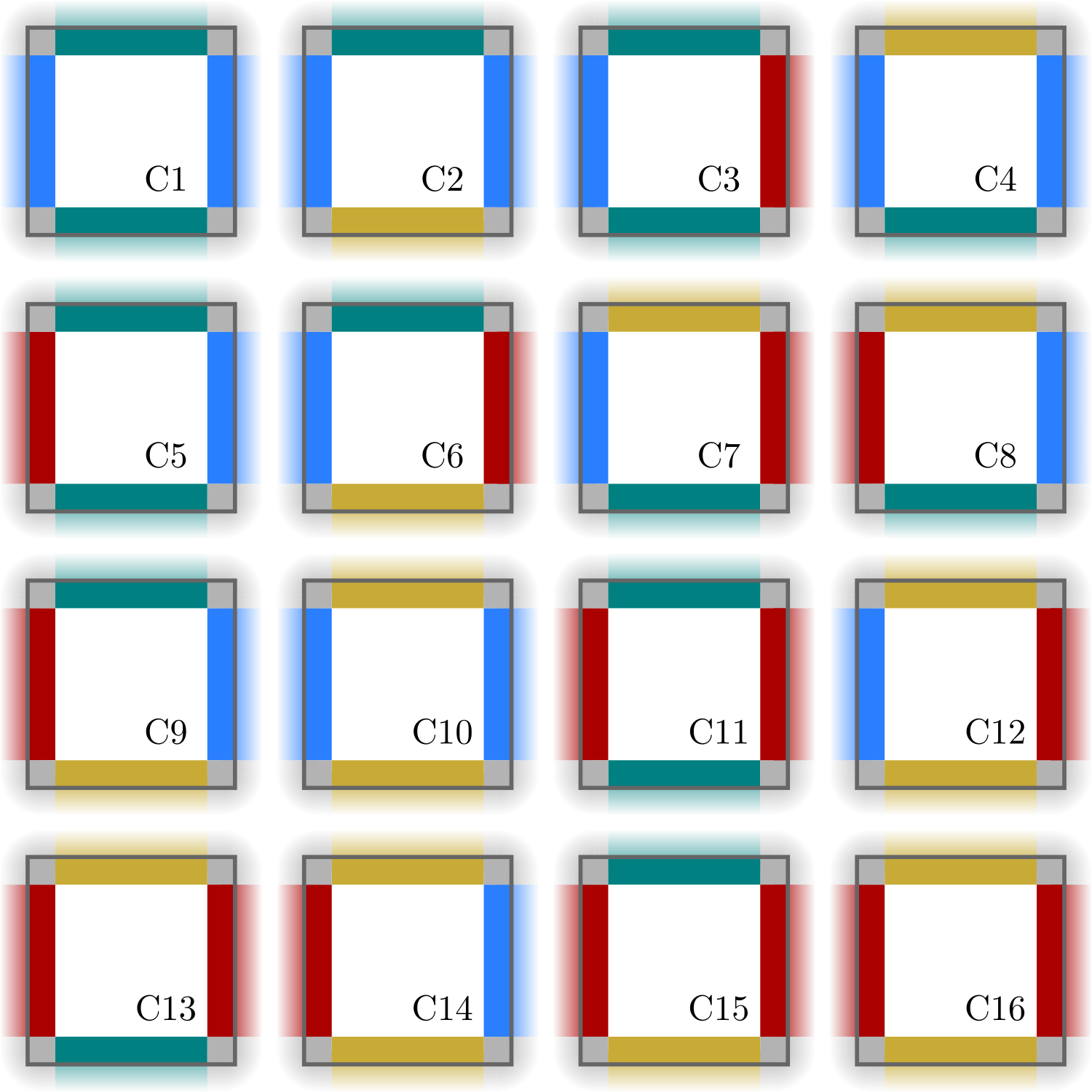} &
		(b) & \includegraphics[width=0.25\textwidth]{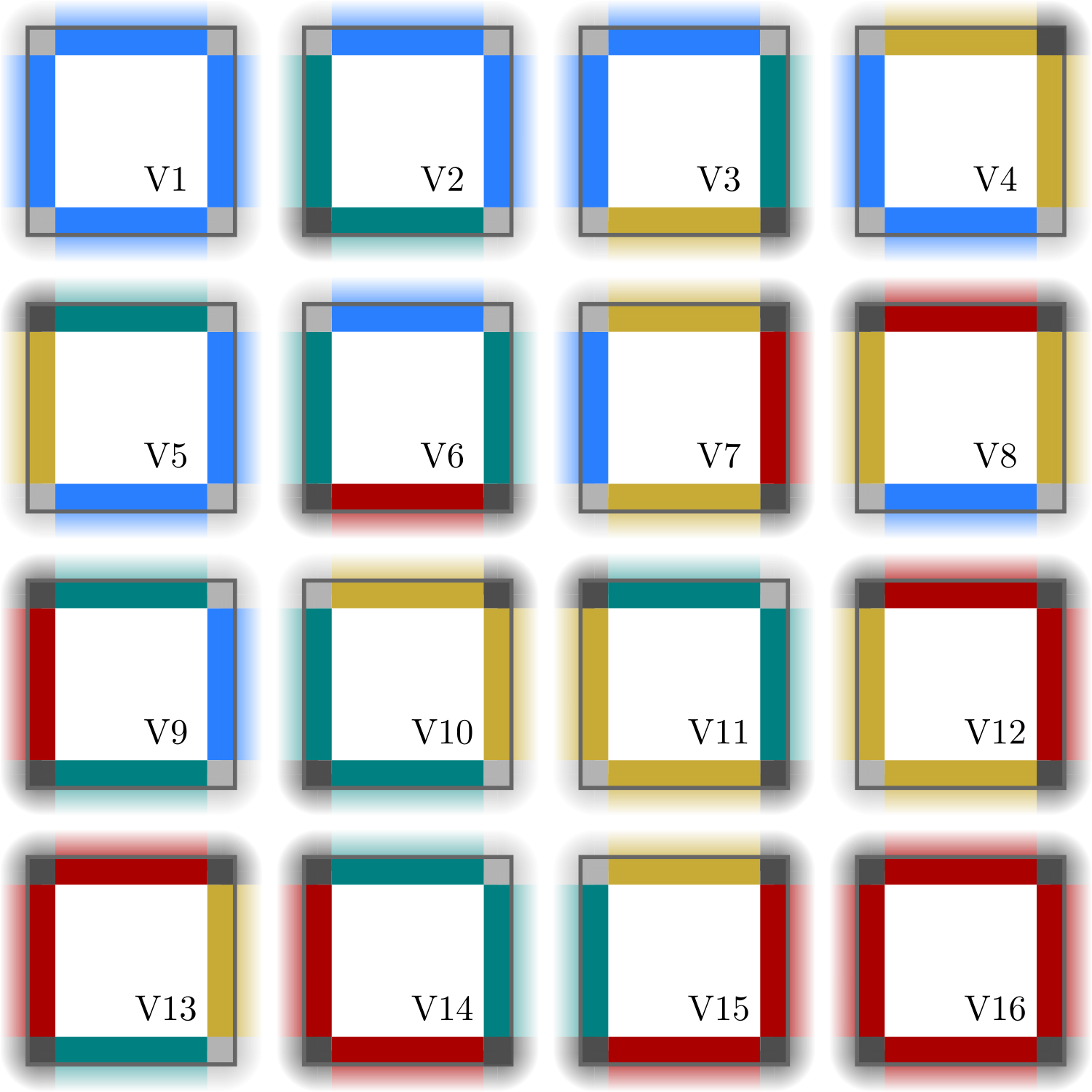} &
		(c) & \includegraphics[width=0.25\textwidth]{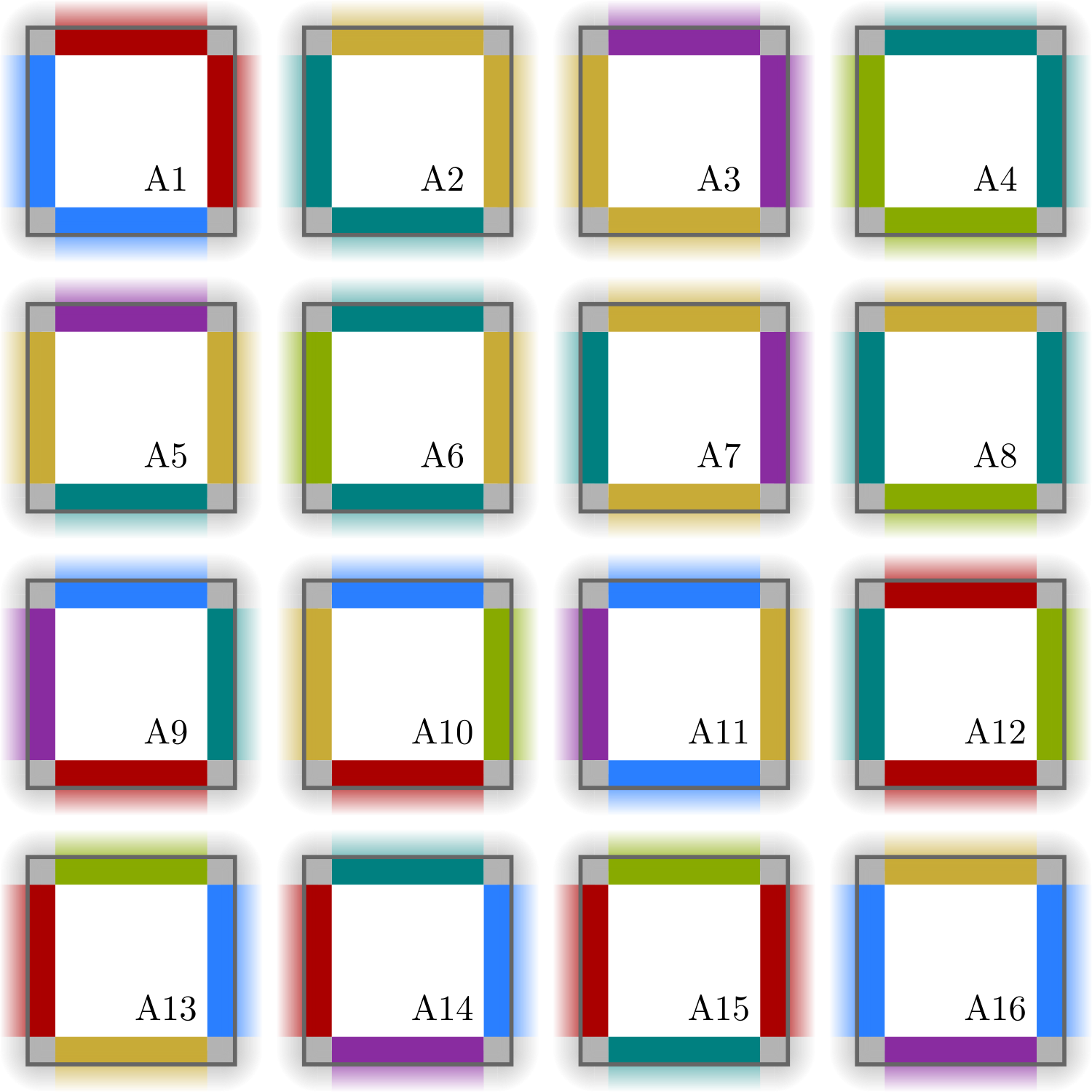}
	\end{tabular}
	\caption{Formal definition of three Wang sets with the equal cardinality of 16 tiles: (a) an edge-based set with two horizontal and vertical codes, (b) a vertex-based set over four horizontal and vertical codes, and (c) the aperiodic Ammann's set~\cite{ammann_aperiodic_1992} over six horizontal and vertical codes. Similar to~\Fref{fig:advertisement}, the compatibility codes playing the role of constraints during the tiling assembly are illustrated with colours assigned to individual edges. The corresponding vertex codes are depicted with light and dark grey corner squares. While in the vertex-based set (b), edge codes were obtained by mapping two vertex codes uniquely to one edge code, the potentially different vertex codes for sets (a) and (b) were identified using the analysis from \Sref{sec:code_analysis} of connectivity graphs shown in \Fref{fig:graphs}.}
	\label{fig:sets}
\end{figure*}

Inspired by applications in computer graphics~\citep{cohen_wang_2003}, we have introduced the formalism of Wang tiles as a suitable generalization of SEPUC representation of heterogeneous materials~\citep{novak_compressing_2012}. The formalism provides a compromise between the SEPUC approach and the use of a microstructure generator for each new realization. The concept of Wang tiles decomposes microstructure generation into an offline phase, which can possibly be computationally intensive, and a nearly instantaneous online phase. In the offline phase, information regarding a heterogeneous microstructure is compressed into a set of smaller domains---Wang tiles---with predefined compatibility constraints on the compressed microstructural geometry. In the online phase, microstructural realizations are assembled from these domains with a fast, linear algorithm that produces stochastic realizations with suppressed periodicity. The merits of the tiling concept for RVE size analyses were demonstrated in~\citep{doskar_jigsaw_2016,doskar_wang_2018}.

Optimization-based approaches developed initially for the SEPUC design can be extended to incorporate generalized periodic boundary conditions and used to generate the morphology of tiles~\citep{novak_compressing_2012,novak_microstructural_2013}. However, the extension amplifies the major weakness of optimization approaches---their computational cost---making them prohibitively expensive for complex three-dimensional models. As a remedy, we have proposed a method motivated by~\citet{cohen_wang_2003} that combines a sample-based approach with quantitative spatial statistics~\citep{doskar_aperiodic_2014}. While this method is by orders of magnitude faster than the optimization approach, it has difficulties handling complex, percolated microstructures such as foam, and produces corrupted ligaments in sample overlaps~\citep{doskar_jigsaw_2016}.

In this paper, we extend Sonon et al.'s method to Wang tiles in order to produce tile-based representations of microstructures intractable by the former methods~\cite{novak_compressing_2012,novak_microstructural_2013,doskar_aperiodic_2014}, see~\Fref{fig:advertisement}.
To this end, in~\Sref{sec:wang_tiling}, we review the fundamentals of the Wang tile concept and discuss, in detail, connectivity and mapping between vertex- and edge/face-based definitions of Wang tiles. Next, we outline Sonon et al.'s method~\citep{sonon_unified_2012,sonon_advanced_2015} and describe modifications necessary to accommodate generalized periodicity, see~\Sref{sec:level_set}. Finally, equipped with the adapted procedure, we illustrate sample outputs of the procedure in two and three dimensions. We also compare three tile sets---two stochastic and a strictly aperiodic one---with the same cardinalities in terms of periodicity artefacts, complementing our previous study~\citep{doskar_aperiodic_2014} that dealt only with the distribution of tile types.

\section{Wang tiles}
\label{sec:wang_tiling}

\emph{Wang tiles} constitute the building blocks of the abstract concept used in this work. Albeit the shape of the tile domains can be any parallelogram (or a parallelepiped in 3D), for simplicity's sake we assume only square (or cubic) domains in the sequel. All tiles from the \emph{tile set} have compatibility codes attributed to their edges (faces), illustrated using colours in Figs. \ref{fig:advertisement} and \ref{fig:sets}. These codes play the role of constraints during an assembly process of tile instances in a \emph{tiling}---a portion of a plane (space) without any voids or overlaps; only tiles with the same codes on the abutting edges (faces) can be placed side by side. In addition, tiles can be neither rotated nor reflected during the assembly. Even though the last two requirements can be eliminated by modifying the definition of compatibility codes~\citep{robinson_undecidability_1971}, we retain them for practical purposes since they preserve the orientation of the microstructure compressed within the tile set. The particular version of an assembly algorithm depends on the type of the tile set, discussed below.

Originally, the concept of Wang tiles was introduced in first-order predicate calculus as a visual surrogate to a decision problem of $\forall\exists\forall$ statements\footnote{A statement containing one existential ($\exists$) and two universal ($\forall$) quantifiers.}~\citep{wang_proving_1961,wang_games_1965}. The initial conjecture that a whole plane can be covered only if a periodic tiling exists~\citep{wang_proving_1961} was disproved shortly after~\citep{berger_undecidability_1966}, triggering the pursuit of the smallest tile set allowing for strictly aperiodic tiling of the plane, see the classical monograph~\citep{grunbaum_tilings_2016} and \citep{jeandel_aperiodic_2015} for historical overviews. This quest for the smallest set appears to be over; supported by an extensive computer-aided search, \citet{jeandel_aperiodic_2015} announced a set of 11 tiles over 4 codes for each edge orientation, stating that no smaller set exists.

\subsection{Stochastic tile sets}

Similar to applications in biology~\citep{rothemund_algorithmic_2004}, physics~\citep{leuzzi_thermodynamics_2000}, and computer graphics~\citep{kopf_recursive_2006,cohen_wang_2003}, we use the concept only in its geometrical interpretation as a suitable formalism describing the mutual compatibility of small domains. Except for the comparison study in~\Sref{sec:set_comparison}, we also limit ourself to the stochastic tile sets introduced by~\citet{cohen_wang_2003}. Besides the fact that deterministic tilings of the aperiodic sets often exhibit locally ordered patterns, the stochastic sets offer higher flexibility in terms of design, i.e. choosing the number of tiles and codes and their distribution within the set\footnote{The limitations of aperiodic sets are especially critical in 3D, because only one aperiodic set of Wang cubes has been published to date~\citep{kari_aperiodic_1995}.}. 

\begin{figure}
	\centering
	\includegraphics[width=0.95\columnwidth]{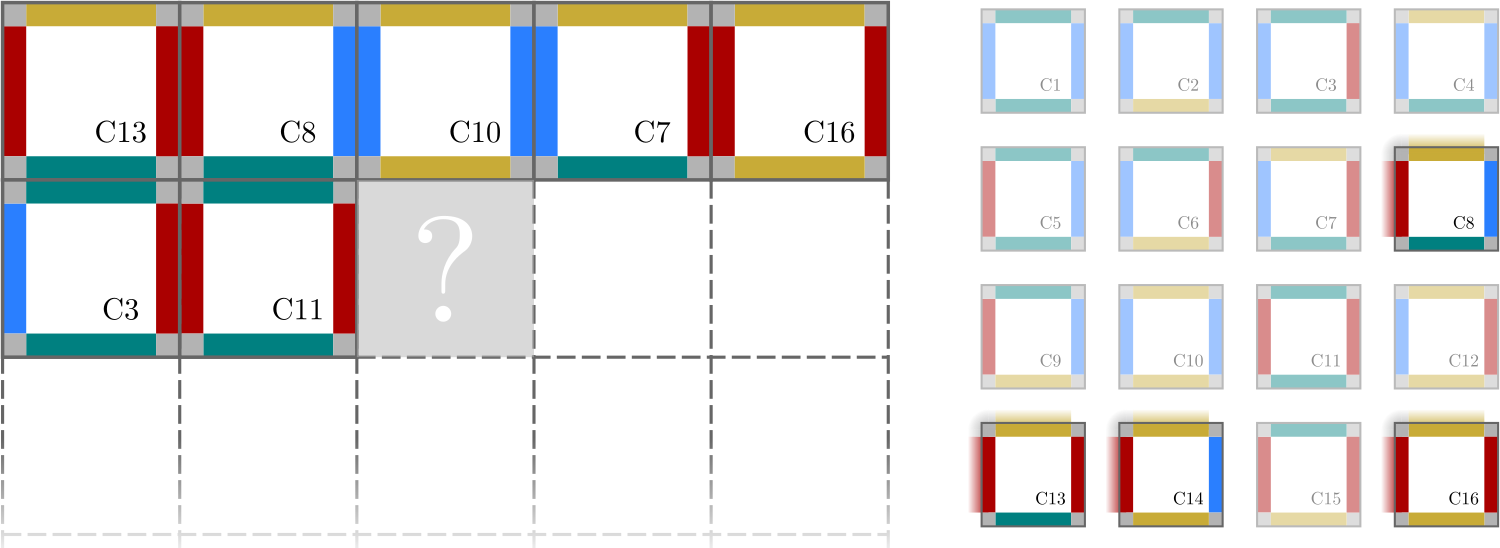}
	\caption{Illustration of a step in the stochastic assembly algorithm~\cite{cohen_wang_2003}. A tiling of a given size is filled sequentially; for each position in a tiling, a final tile is chosen from candidate tiles that are filtered out from a given tile set based on the compatibility constraints posed by the previously placed tiles. The candidate tiles for the particular position denoted with the question mark are highlighted in the tile set shown on the right.}
	\label{fig:assembly_illustration}
\end{figure}
The assembly algorithm for stochastic sets works sequentially: an initially empty grid is traversed in a scanline~\citep{lagae_alternative_2006} way; at each grid node, possible candidates compatible with previously placed tiles are identified in the set and one candidate tile is randomly chosen and placed; see the illustration in~\Fref{fig:assembly_illustration}. Thus, the only requirement for the set design is that there is at least one tile (but optimally two or more to preserve the stochastic nature of the assembly) for every possible combination of codes on the upper and left-hand edges. 
In principle, the random choice from the subset can be replaced with an informed selection preferring e.g. different phase volume fractions in different regions of a domain according to a pre-generated Gaussian random field. This modification allows for correlations in the microstructure at length scales larger than the tile size. For the sake of brevity and without loss of generality, however, we use only the standard stochastic algorithm in this work.

The tiling concept's ability to generate naturally looking patterns from a limited amount of samples---which proved highly appealing in computer graphics~\citep{cohen_wang_2003,kopf_recursive_2006}---stems from the reduced periodicity in tiling assemblies. 
This feature complies well with our ambition to replace a SEPUC-based description of random heterogeneous materials with its generalization that would allow for fast synthesis of stochastic microstructural samples or microstructure geometries for entire macroscopic domains.

The design of a compressed microstructural representation consists of two steps. First, the cardinality of the tile set and a particular distribution of edge (face) codes is chosen. This controls the frequency of tile occurrence in a tiling. Next, tile interiors are designed such that (i) the generated microstructure is continuous across the compatible edges and (ii) assembled tilings resemble the desired microstructure.
Note that the latter requirement is not imposed on individual tiles. On the contrary, variability in the compressed representation is the main merit of the tiling concept. Tile interiors together with edge compatibility carry local microstructural characteristics, while the fluctuations over distances larger than the tile size are facilitated via the tile assembly algorithm.

\subsection{Edge- vs. vertex-based tile definitions}

Complementary to the edge-based specification introduced above, Wang tiles can be defined using vertex codes. Unlike the standard edge-based definition, the vertex-based definition allows for direct control of tile states across vertices, preventing pronounced repetitiveness when a visually distinctive microstructural feature falls into the vertex region~\citep{cohen_wang_2003}. To avoid this \enquote{corner problem}, \citet{cohen_wang_2003} proposed marking the tile vertices with an additional set of codes, essentially overlaying two tile set definitions. Subsequently, \citet{lagae_alternative_2006} retained only vertex codes, reporting superior spectral properties for assembled patterns compared to edge-based sets with the same cardinality.

Albeit the two definitions yield the same cardinality of tiles ($C^4$) in the complete set, i.e. the set that contains all combinations of $C$~codes, they differ when it comes to minimal stochastic sets, i.e. sets that contain at least two tiles for each admissible code combination of already placed neighbouring tiles ($2C^2$ for edge-based sets vs. $2C^3$ in the case of vertex-based ones). The difference is even more pronounced in three dimensions: a full face-defined set contains $C^6$ cubes compared to $C^8$ cubes in a vertex-defined set. Theoretically, Wang cubes can be also defined with edge codes in three dimensions; however, cardinality exceeds both the face- and vertex-based definitions without any known benefits.

Mapping vertex-defined tiles\footnote{The same procedure holds also for Wang cubes, where four vertices define one face code.} to the original definition is straightforward: two vertex codes define one edge code. Hence all the above-mentioned design and assembly procedures directly apply to the vertex-based definition as well. In fact, the mapping provides an effortless way to produce a minimal stochastic edge-defined set with equal occurrence probability of each tile, the set characteristics sought for in~\cite{novak_microstructural_2013}. On the other hand, because the mapping is injective only, it is generally not possible to map an edge-defined set to a vertex-based one.

\subsection{Vertex analysis of a tile set}
\label{sec:code_analysis}
\begin{figure}[ht!]
	\centering
	\includegraphics[width=\columnwidth]{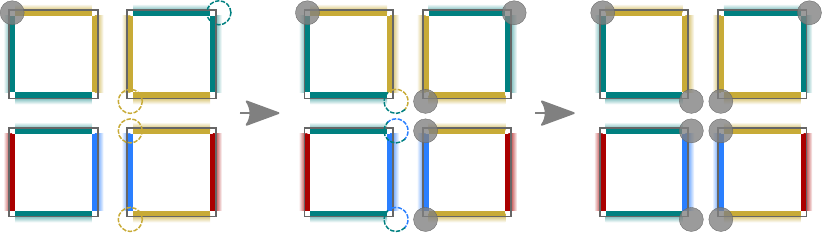}
	\caption{Illustration of the generalized periodicity effect on the particle copy process. The three steps shown illustrate how a particle initially placed at the top left corner of the first tile (grey particle on the left-hand side of the figure) is sequentially copied to other loci following the edge codes it intersects. Dashed particle outlines loci where the particle will be copied in the next step.}
	\label{fig:copy}
\end{figure}
\newcolumntype{S}{>{\arraybackslash} m{.25\linewidth} }
\begin{figure*}[ht!]
\centering
\setlength{\tabcolsep}{4pt}	
\begin{tabular}{cp{0.25\textwidth}cp{0.28\textwidth}cp{0.28\textwidth}}
	(a) & \includegraphics[width=0.25\textwidth]{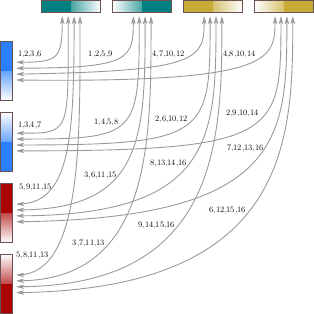} &
	(b) & \includegraphics[width=0.25\textwidth]{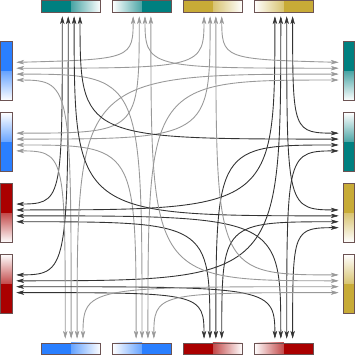} &
	(c) & \includegraphics[width=0.25\textwidth]{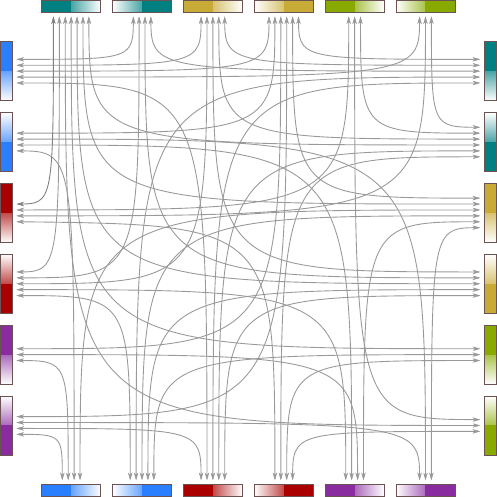}
\end{tabular}
\caption{Comparison of graphs pertinent to the tile sets from~\Fref{fig:sets}, identifying a potential vertex-based definition of the sets. Each graph node corresponds to a code on a half of a tile edge (either left/right for horizontal codes, or top/bottom for vertical codes), and graph arcs correspond to tile vertices. As there are multiple vertices---and hence parallel arcs---connecting the same nodes, the graphs should be drawn as a multigraph. However, multiple arcs are collapsed into one for the sake of brevity; only in (a) multiplicity is indicated by tile indices attached to each arc. Independent sub-graphs indicating the vertex character of the set are shown in distinct light and dark grey, which correspond to the vertex codes in~\Fref{fig:sets}.}
\label{fig:graphs}
\end{figure*}

Here, we outline a tile set analysis capable of revealing the underlying vertex definition, if it exists. The motivation behind this analysis is the pragmatic question which arises when implementing generalized periodicity: If a particle intersects a tile boundary, what other tiles should it be copied to? 
While the answer is straightforward for a particle intersecting a tile edge, vertex overlap is more involved. A particle overlapping a vertex is carried to other tiles by both the horizontal and vertical edges. Because the particle overlapped a vertex in the original tile, its images overlap vertices of the other tiles as well. Consequently, these images are propagated further by the newly affected edge codes, see~\Fref{fig:copy}. Whether the particle is eventually copied to all vertices or appears only in a selected subset depends on the allocation of codes to individual tiles.

Assume an undirected graph\footnote{Instead of the standard vertex-edge nomenclature for a graph, we use node and arc terms in order to avoid confusion with similar geometrical notions related to Wang tiles.} where each node represents the code on a particular half of an edge (either top/bottom for vertical or left/right for horizontal codes, resulting in two occurrences of each edge code) and each arc corresponds to a tile vertex. The graph is by definition bipartite, because each vertex connects horizontal and vertical codes, and represents a multigraph since there are usually more vertices with the same adjacent edge codes in the tile set. 
To answer the aforementioned question, we identify connected components of the graph using the Depth-First Search algorithm~\cite[Section 18.2]{sedgewick_algorithms_2002}. 
Each independent sub-graph then corresponds to a distinct vertex code and the arcs pertaining to the sub-graph determine the vertices to which a vertex-overlapping particle will be propagated.
See~\Fref{fig:graphs} for a comparison of three graphs pertinent to the tile sets depicted in~\Fref{fig:sets}. The subgraphs (if present) are plotted in distinct grey colours, which correspond to the colours of vertex codes shown in~\Fref{fig:sets}.

In three dimensions, a set of Wang cubes can be analysed analogously with only a minor modification: a graph node represents one corner of a cube face. Thus, each face code appears four times in the graph. In addition, cube edges with the same direction vector must be also classified, addressing the situation when a particle intersects a cube edge. The classification follows the exact same procedure as vertex identification in two dimensions, neglecting the codes on faces perpendicular to the analysed edges.

\section{Tile design using level-set functions}
\label{sec:level_set}

As mentioned in \Sref{sec:state_of_the_art}, Random Sequential Adsorption (RSA)\footnote{Sometimes, the term Random Sequential Addition, e.g.~\citep{talbot_random_1991}, is used interchangeably.} is one of the most frequent algorithms for generating particle packings and microstructural geometries. It follows the simple idea of throwing particles of an arbitrary shape into a given domain and keeping those that do not overlap with previously placed ones. In its original setting, however, the algorithm poses a critical drawback for higher volume fractions: the success rate for accepting the particle position rapidly decays in later stages, because the majority of the randomly selected positions collide with the already placed particles. Optimally, the remedy would be to sample the new particle position directly from a domain that is certified to result in a non-overlapping composition.

Several such remedies have been listed in~\Sref{sec:state_of_the_art}. Here, we recall the approach of Sonon et al., because we have extended their methodology to Wang tiles. Sonon et al.~\cite{sonon_unified_2012,sonon_level-set_2013,sonon_advanced_2015,massart_level_2017} adopted the implicit, level-set-based description of a microstructural geometry and demonstrated that it enables---in addition to the desired sampling from a valid domain loci---generating complex microstructural geometries (such as open and closed foams) in a unified framework, using suitable morphing operations. Moreover, both features are independent: if a particle packing is the desired output, the morphing operations can be neglected. On the other hand, any distribution of points or particles obtained by different packing methods can serve as the input for the morphing operations. 

In the the remainder of \Sref{sec:level_set}, we revise individual steps from the Sonon et al.'s framework and introduce necessary modifications facilitating the compatibility constraints arising in the concept of Wang tiles. For the sake of clarity, all procedures are presented in the two-dimensional setting with comments about three-dimensional problems when necessary. \Sref{sec:results} then presents both two- and three-dimensional results.

\subsection{Generating particle packings}
\label{sec:packing}

The original Sonon et al.'s approach~\citep{sonon_unified_2012} rests on describing the geometry of particle $\particle$ with a level-set function $\LSp$, which gives the signed distance of point $\x$ from the nearest particle boundary $\particleBoundary$ (with negative values inside the particle), such that 
\begin{figure}
	\centering
	\includegraphics[width=0.9\columnwidth]{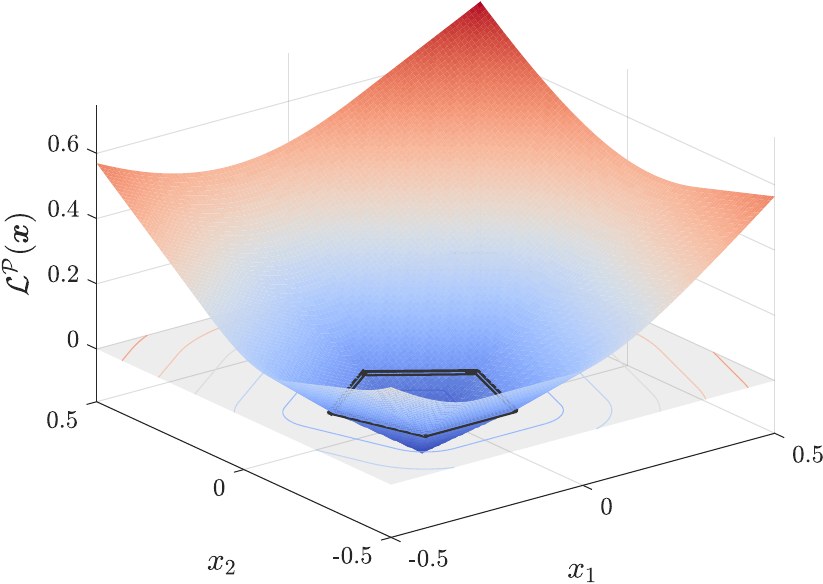}
	\caption{Implicit representation of particle geometry with the level-set function $\LSp$. The boundary of particle $\mathcal{P}$ is obtained as a zero-value contour of $\LSp$ (plotted in a black line).}
	\label{fig:LS_illustration}
\end{figure}
\begin{equation}
	\LSp\atx = \min_{\y \in \particleBoundary} d\left( \x, \y \right),
\end{equation}
with the signed distance function $d\left( \x, \y \right)$ given by
\begin{equation}
	d(\x,\y) = 
	\begin{cases}
		-\!\norm{\x-\y} &\quad \text{if} ~\x \in \particle, \\
		\hphantom{-}\!\norm{\x-\y} &\quad \text{otherwise}\,. \\
	\end{cases}
\end{equation}
Consequently, the zero iso-line of $\LSp$ represents the particle boundary $\particleBoundary$, as depicted in~\Fref{fig:LS_illustration}. Note that unlike many applications of the level-set method where the implicit geometry description evolves in time according to the Hamilton-Jacobi equation, see~\cite[and references therein]{osher_level_2003}, $\LSp$ is computed for the given geometry of a particle and remains fixed.

\begin{figure}
	\setlength{\tabcolsep}{2pt}
	\begin{tabular}{ccc}
		\includegraphics[height=3.75cm]{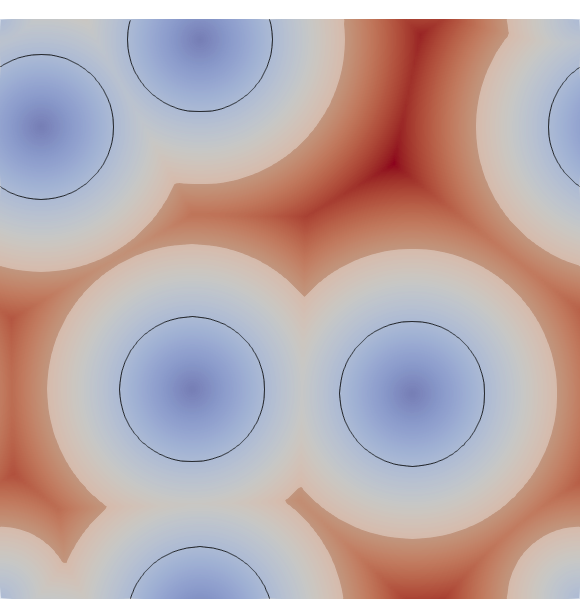} & \includegraphics[height=3.75cm]{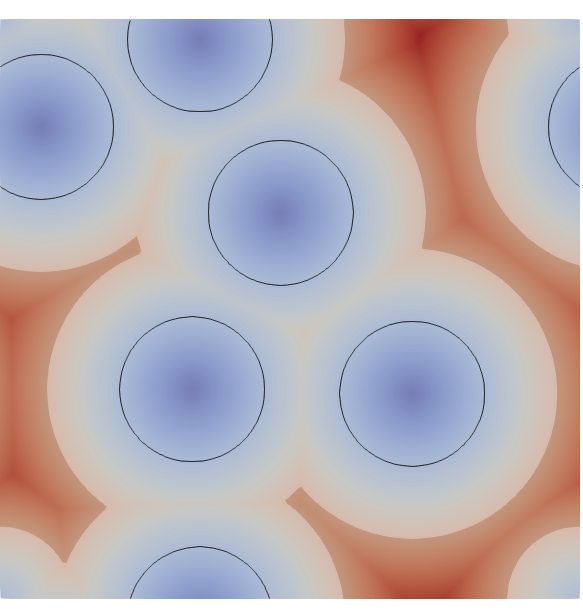} & \includegraphics[height=3.75cm]{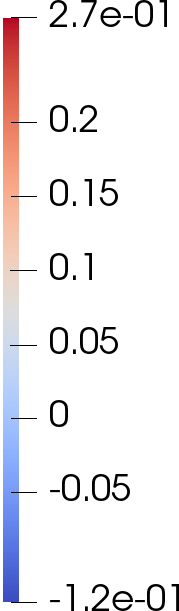}\\
		(a) & (b)
	\end{tabular}
	\caption{Two consecutive steps of the level-set based RSA algorithm for a Periodic Unit Cell: (a) level-set field $\LSd$ with previously placed particles (values of $\LSd$ outside the admissible sampling domain $\admissibleDomain$ are displayed as translucent) at the $n$th step; (b) updated state of $\LSd$ and $\admissibleDomain$ after placing a new particle at step $n+1$.}
	\label{fig:step_in_LSRSA}
\end{figure}

Assume for now that we have domain $\domain$ that contains set $\mathcal{R}$ of already placed particles and their geometry is encoded in a single level-set field $\LSd$, 
\begin{equation}
	\LSd\atx = \min_{\particle \in \mathcal{R}} \LSp\atx\,, \quad \forall \x \in \domain \,;
\end{equation}
see~\Fref{fig:step_in_LSRSA} for an illustration.
Knowing the radius $\radius$ of the smallest circumscribed circle as the only characterization of the new particle, overlaps with the existing particles can be readily prevented by sampling its centre from the domain
\begin{equation}
	\admissibleDomain = \left\{\x\in\domain \mid \LSd\atx \geq \radius \right\}.
	\label{eq:admissible_domain}
\end{equation} 
After placing the new particle $\particle$, the level-set field $\LSd$ is updated with $\LSp$,
\begin{equation}
	\LSd\atx = \min \left( \LSd\atx, \LSp\atx  \right), \quad \forall \x \in \domain \,.
	\label{eq:ls_update}
\end{equation}
This procedure repeats until either a pre-defined stopping criterion (e.g. a desired volume fraction) is met or $\admissibleDomain = \emptyset$. 

To increase the volume fraction that can be achieved with this procedure, two additional restrictions can be posed on the admissible subdomain \admissibleDomain. 
First, a maximal distance $\rho$ from existing particle boundaries can be added to~\Eref{eq:admissible_domain}, preventing too large gaps between particles. (If requested, a minimal distance $\kappa$ between two particles can be further enforced as well.)
Second, to prevent locally jammed states with large interparticle gaps, an additional field $\LSdi{\text{II}}$ storing the shortest distance to the surface of the second nearest particle can be included in~\Eref{eq:admissible_domain} and limited from above with $\sigma$, see \citep{sonon_unified_2012} for additional details. 
With all these constraints in action, \Eref{eq:admissible_domain} takes the form
\begin{equation}
	\admissibleDomain = \left\{\x\in\domain \mid \radius + \kappa \leq \LSd\atx \leq \radius + \rho,\,\LSdi{\text{II}} \leq \radius + \sigma \right\}.
	\label{eq:admissible_domain_contraints}
\end{equation} 

Enforcing periodicity of the packing is also straightforward: upon placement of a new particle, the domain level-set field is also updated with the particle's periodic images (eight in 2D; recall the grid in~\Fref{fig:copy_preventer}, and 26 in 3D).

\subsubsection{Extension to Wang tiles}
\label{sec:extension_tiles}
\begin{figure}
	\centering
	\includegraphics[width=0.65\columnwidth]{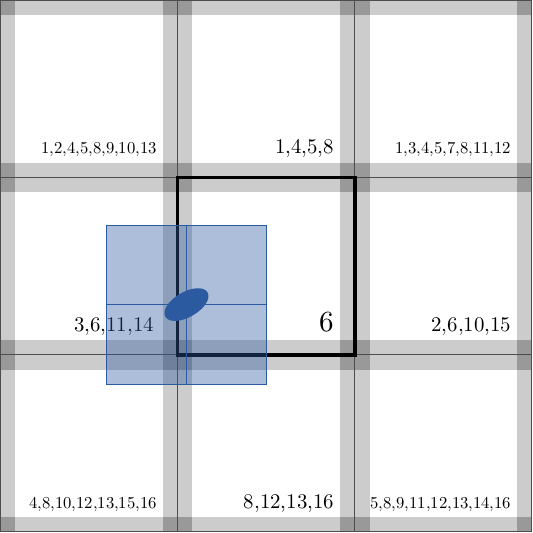}
	\caption{Neighbourhood grid identified for each tile in a tile set (here shown for tile V6 from \Fref{fig:sets}b) containing codes of tiles that can occur at given positions. Light and dark grey regions depict individual entities in the bookkeeping structure, which prevents calculations. (During the update procedure described in the text, the blue particle can be accounted for either in tile 6 at the centre, or from tiles 3, 6, 11, or 14 at the first column, second-row position. With the bookkeeping structure considered, only one update is performed.) The blue square illustrates the use of acceleration via a precomputed patch; values of $\LSp{}$ are computed only once and used for all particle occurrences during an update. If requested, values outside the patch are computed in a regular way.}
	\label{fig:copy_preventer}
\end{figure}

Compared to generating PUCs, the additional compatibility constraints among individual Wang tiles necessitate several modifications in the original Sonot et al.'s algorithm to generate continuous microstructural morphology of a tile set. 
For the Poisson discs distribution, \citet{lagae_poisson_2006} modified the Dart Throwing algorithm by partitioning tile domains into separate regions pertinent to vertices, edges, faces, and tile interiors, and filling the regions sequentially, starting with the vertex regions. Here we present a different approach better suited to packing particles of different sizes and shapes that processes all tiles simultaneously and does not require a priori partitioning of tile domains.

As $\LSd$ we use a level-set field $\LSs{} = \bigcup_{\tile{}\in\tileset} \LS^\tile{}$ defined on a regular grid for each tile $\tile{}$ in the tile set $\tileset$. 
For each tile in a pre-processing stage, we identify potential neighbouring tiles and store their indices in a grid ($3\times3$ or $3\times3\times3$ in two or three dimensions, respectively), see~\Fref{fig:copy_preventer}, which will be used later in the updating phase.

For further description, we define a \emph{copy-inducer} as a geometrical entity that is responsible for inducing particle copies to other tiles. Hence, a copy-inducer can be either a vertex, edge, face, or formally the tile itself and is related to a code identified in~\Sref{sec:code_analysis}. The particular type of the copy-inducer stems from particle's intersections with a tile boundary (if a particle does not intersect a boundary, the tile is the copy-inducer, indicating that no copy is necessary. If a particle intersects one edge, the edge is the copy-inducer, and so on).

\paragraph{Modification 1: Update}
The main structure of the modified algorithm follows the original one.
After sampling a particle location from $\admissibleDomain \subseteq \tileset$, which yields tile $\tile{i}$ and centre $\x_c^\particle \in \tile{i}$ where the new particle will be placed, the particle's intersections with the boundary of $\tile{i}$ are determined, and---based on the intersections---a copy-inducer is identified. Images of the particle are subsequently copied to the relevant tiles following the matching codes on the corresponding copy-inducers. A level-set field of the whole set is then updated tile-wise as described below.

For each tile, we sequentially loop over all tile positions in the neighbours grid shown in~\Fref{fig:copy_preventer} (including the centre position as well) and check each of the potential tiles at that position for a new particles added in the current step. If any, the central tile's level-set field is updated according to~\Eref{eq:ls_update}. Clearly, a new particle can appear several times in the same place. To avoid duplicated calculations, we set up a bookkeeping structure that records which copy-inducer in the neighbouring grid was already used during the update and run only the unperformed updates, see~\Fref{fig:copy_preventer}.

Two implementation approaches are possible at this stage: (i) $\LSp$ is computed anew for each tile and position in the bookkeeping structure, or (ii) provided that $\LSs{}$ is represented with discrete values on a regular grid, $\LSp$ is precomputed for an auxiliary domain/patch aligned with the particle centre before any update and this field is then reused for all particle instances. 
Moreover, we combine both approaches with the pre-screening acceleration proposed in~\cite{sonon_unified_2012} that resorts to computing potentially demanding surface-distance functions only when the $\LS$ value of a circumscribed circle/sphere---which is fast and easy to calculate---is less than the actual $\LSd$ value for a given $\x$. 

While the pre-screening acceleration is significant in all cases, the benefits of a precomputed patch depend mainly on the patch size relative to the particle circumscribed radius, the $\LSs{}$ field resolution, and the cardinality of the tile set. Especially in the later stages of the algorithm, when only relatively small parts of $\LSs{}$ are updated, pre-computing a patch for a particle that is not copied to other tiles might cause unnecessary overhead. Therefore, we use the patch approach only for particles with either vertex, edge, or face copy-inducers.

For the three-dimensional set of 16 cubes used in~\Sref{sec:3Ddemo} with ellipsoidal particles of circumscribed radius $\radius = 0.15$, the patch pre-calculation delivered 10\% saving in computational time on average.
Note also that the multi-query problem of finding the shortest distance to a particle boundary from a set of points is trivially parallelizable, changing the trade-off between patch pre-computations and direct calculations with available threads.

When all tiles are updated, the algorithm proceeds with identifying $\admissibleDomain$ for the next particle.

\begin{figure}[t]
	\centering
	\begin{tabular}{cc}
		\multicolumn{2}{c}{\includegraphics[width=0.8\columnwidth]{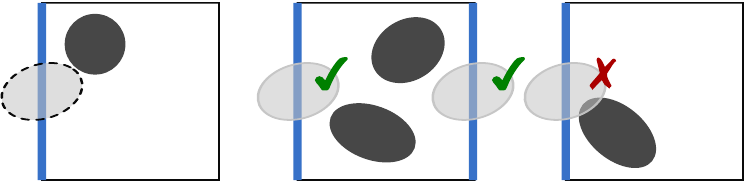}} \\
		\multicolumn{2}{c}{(a)} \\
		\includegraphics[width=0.425\columnwidth]{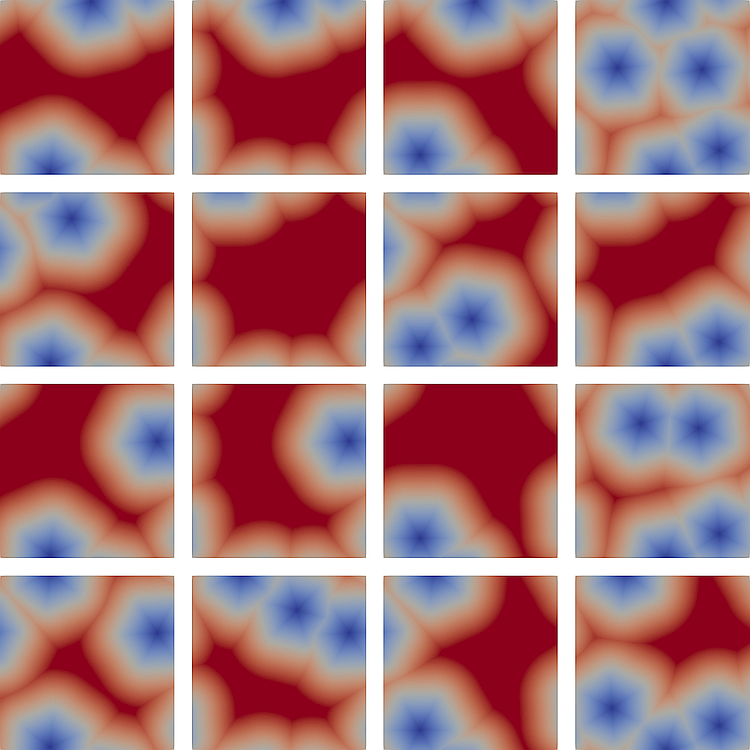} & \includegraphics[width=0.425\columnwidth]{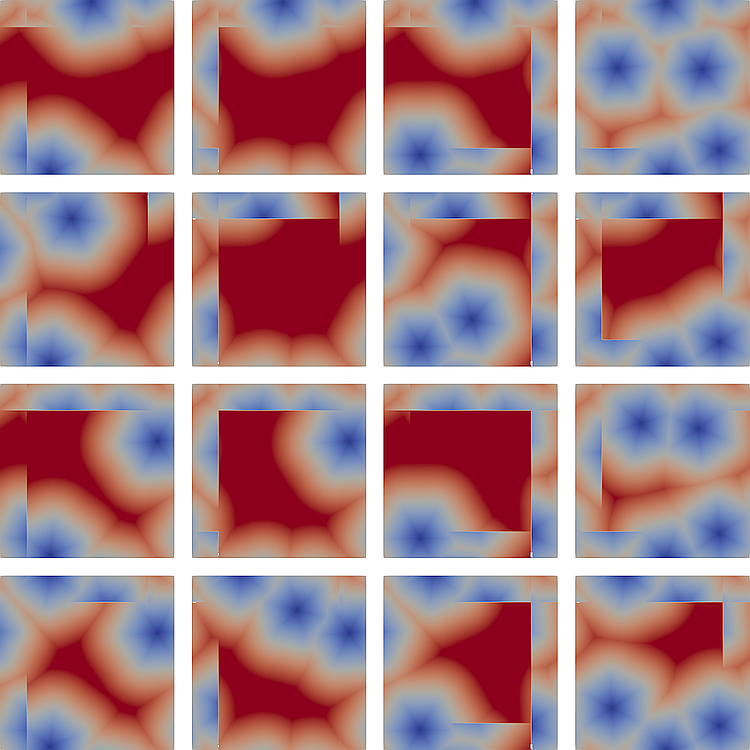} \\
		(b) & (c)
	\end{tabular}
	\caption{ (a) Illustrations of particle intersections that can happen due to copying particle images without considering artificially updated level-set field $\widetilde{\LS}^{\tileset}$ in the definition of $\admissibleDomain$. The problematic overlaps occur only at inducer places (highlighted in blue) with the same orientation as the particle copy-inducer (dashed outlined grey particle on the left-hand side). Figures (b) and (c) depict the original $\LSsSimple$ and its modified version $\widetilde{\LS}^{\tileset}$ that communicates necessary data across boundary regions to prevent intersections from (a).}
	\label{fig:overlap_illustration}
\end{figure}

\paragraph{Modification 2: Artificial level-set field}
Definition of the admissible domain $\admissibleDomain$ must also be modified for Wang tiles; particle centres cannot be sampled directly from $\admissibleDomain$ obtained from $\LSs{}$. The reason for this is that the near-boundary parts of each tile must be informed about the interiors of related tiles in order to prevent insertions similar to the one depicted in~\Fref{fig:overlap_illustration}a, where a copied instance of a newly placed particle collides with interior particle(s) of other tiles. 
The problem arises only when a particle image is added to a copy-inducing entity with the same orientation as the actual inducer (left-hand side edges in \Fref{fig:overlap_illustration}a); particles copied to the opposite copy-inducers automatically meet the non-overlap requirement thanks to the first update modification described above.

To avoid this problem, we replace $\LSs{}$ in \Eref{eq:admissible_domain} or \Eref{eq:admissible_domain_contraints}\footnote{Recall that the tile set domain $\LSs{}$ substitutes a generic $\LSd$ in both equations from \Sref{sec:extension_tiles} onwards.} with a modified field $\widetilde{\LS}^{\tileset}$ which is equal to $\LSs{}$ except for boundary regions, whose width is dictated by the radius $\radius$ of the circle/sphere circumscribed to the current particle. Each boundary region is then constructed as a point-wise minimum over all regions related to the same copy-inducers with the same code and the same orientation; see~\Fref{fig:overlap_illustration}c. This construction propagates the close-boundary state across relevant tiles and thus prevents overlapping insertions similar to the one depicted in~\Fref{fig:overlap_illustration}a. Note that a similar modification is not required for $\LSdi{\text{II}}$ because all information necessary to prevent the collisions is already contained in $\widetilde{\LS}^{\tileset}$.

\paragraph{Modification 3: Breaking regular grid} 
The level-set field $\LSs{}$ is typically implemented on a regular grid and the particle centres $\x_c^\particle$ thus end up aligned with the grid. To break this artificial ordering, after sampling a new particle centre, we check whether the surrounding points belong to $\admissibleDomain$. The belonging points then define quadrants/octants in which the particle can be moved freely. We generate a random shift within these quadrants/octants and update the particle centre accordingly. However, this modification is possible only if the grid spacing is sufficiently small compared to the particle size; otherwise, the implicitly assumed linear approximation of $\LSs{}$ is inaccurate and might result in particle intersections at intermediate locations.

The modifications described above are sufficient for extending the original Sonon et al.'s level-set-based packing algorithm to the concept of Wang tiles. Note that all control variables, e.g. particle radius $\radius$ or minimal and maximal distances $\rho$ and $\sigma$, can change while the algorithm operates. For instance, to achieve denser packings for multi-modal particle size distributions, it is preferential to start placing large particles first and sequentially proceed to the smaller ones.

Nonetheless, our extension inherits the weakness of the original algorithm when it comes to particles whose shapes are significantly different than their circumscribed circle/sphere, e.g. prolonged ellipsoids. In particular, the more the particle volume deviates from the volume of the circumscribed sphere, the less dense the final packing can be.
While the geometry of the already placed particles is described exactly with $\LSs{}$ (up to inaccuracies due to grid spacing in $\x$), the newly placed particle is represented only approximately during selection of a centre, yielding a pessimistic estimate of mutual intersections. The potential remedy is to translate the particle centre right after sampling according to some heuristic rule aimed at producing denser packings; however, we leave this issue unaddressed in this paper.

\subsection{Morphing operations}
\label{sec:morphing}

The acceleration of RSA by adopting the level-set approach is attractive on its own; however, the main appeal of the Sonon et al.'s level-set framework is the elegance with which complex microstructures can be generated.
As the simplest example, adding a constant $\gamma$ to $\LSs{}$ enables (i) fine-tuning of particle volume fractions, (ii) smoothing of sharp corners of polyhedral particles, and (iii) particle coating, e.g. for defining the Interfacial Transition Zone.

Combinations of $\LSs{}$ and $\LSs{2}$ permit rendering interparticle bridges and controlled Vorono\"{i}-like tessellations; see~\citep{sonon_unified_2012}, where these morphing operations were used to mimic microstructural geometry of clay/sand mixed soils and irregular masonry.
Considering also the shortest distance to the boundary of the third nearest particle $\LSs{3}$, \citet{sonon_advanced_2015} demonstrated that their framework can produce highly adjustable models of foam-like microstructures, including features such as a smooth transition from open to closed foams, concavity of foam ligaments, their coatings and hollow interiors, and variable thickness of foam ligaments and cell walls.

As in \Sref{sec:packing}, we present only the essentials of generating foam-like microstructures and introduce necessary modifications. The reader is referred to~\citep[Section 4]{sonon_unified_2012} and \citep[Sections 3 and 4]{sonon_advanced_2015} for details regarding the above-mentioned local adjustments, since they remain unchanged.

Assume we have at our disposal all three fields $\LSs{}$, $\LSs{2}$, and $\LSs{3}$, containing the shortest distance to the boundary of three nearest particles (either computed a posteriori for a given particle assembly or already tracked during the packing algorithm).

\begin{figure}
	\centering
	\begin{tabular}{cc}
		\includegraphics[width=0.45\columnwidth]{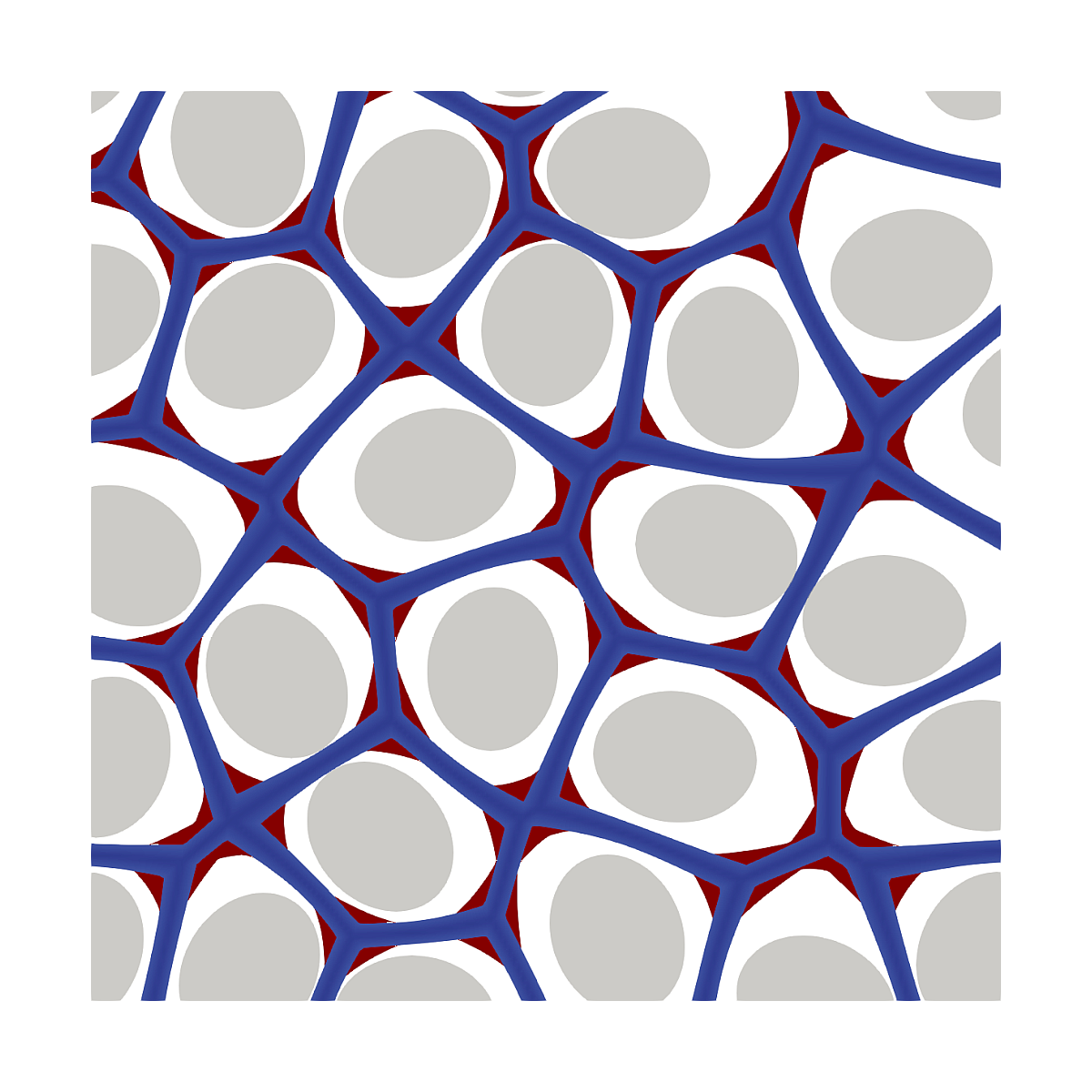} & \includegraphics[width=0.45\columnwidth]{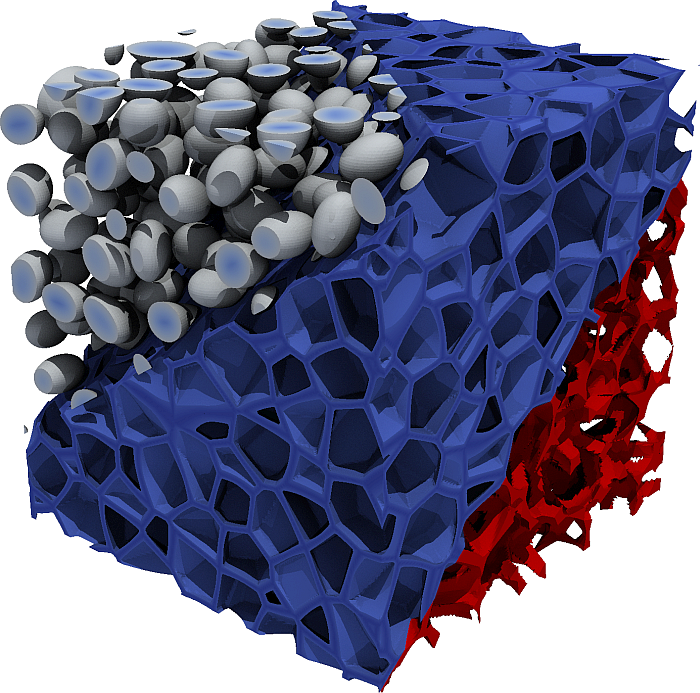}\\
		(a) & (b)
	\end{tabular}
	\caption{Influence of morphing operations on resulting (a) two- and (b) three-dimensional geometry. From the original particle distribution (depicted with light grey), the closed-cell, foam-like geometry (shown in blue) is obtained by~\Eref{eq:closed_foam} and the open-like features (plotted in red) follow from~\Eref{eq:open_foam}.}
	\label{fig:morphing}
\end{figure}

In a manner similar to classical Vorono\"{i} tessellation, where a domain is partitioned by boundaries that have the same distance to the two closest seeds, thresholding the modified difference
\begin{equation}
	\mathcal{F}_{\textrm{c}} =  \left(\LSs{2} - \LSs{}\right) + t_{\textrm{c}}
	\label{eq:closed_foam}
\end{equation}
yields a closed-cell, foam-like geometry with $t_{\textrm{c}}$ indirectly controlling the thickness of foam cell walls. 
In the same spirit, the centrelines of the open-foam ligaments can be viewed as loci with the same distances to the boundary of three nearest particles; hence thresholding
\begin{equation}
	\mathcal{F}_{\textrm{o}} = \left(\tfrac{1}{2}(\LSs{3} + \LSs{2}) - \LSs{}\right) + t_{\textrm{o}}\,,
	\label{eq:open_foam}
\end{equation}
produces open-foam ligaments with $t_{\textrm{o}}$ governing their thickness. 
Finally, combining Eqs. (\ref{eq:closed_foam}) and (\ref{eq:open_foam})
\begin{equation}
	\mathcal{F}\atx = \min\left(\mathcal{F}_{\textrm{c}}\atx,\mathcal{F}_{\textrm{o}}\atx\right), \quad \forall \x \in \tileset \,,
	\label{eq:foam_composition}
\end{equation}
allows for more realistic geometries with material concentration at wall intersections, see~\Fref{fig:morphing}.

\begin{figure}[ht!]
	\centering
	\setlength{\tabcolsep}{3pt}
	\begin{tabular}{ccc}
		\includegraphics[width=0.3\columnwidth]{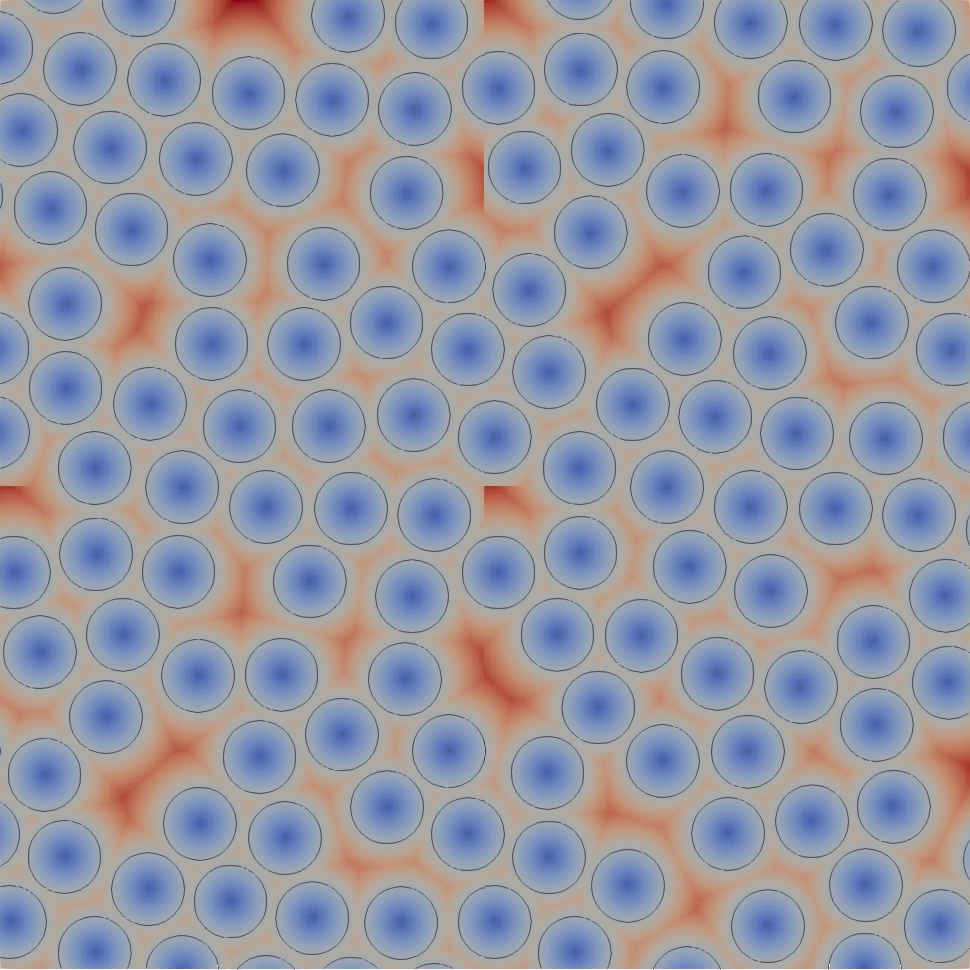} & \includegraphics[width=0.3\columnwidth]{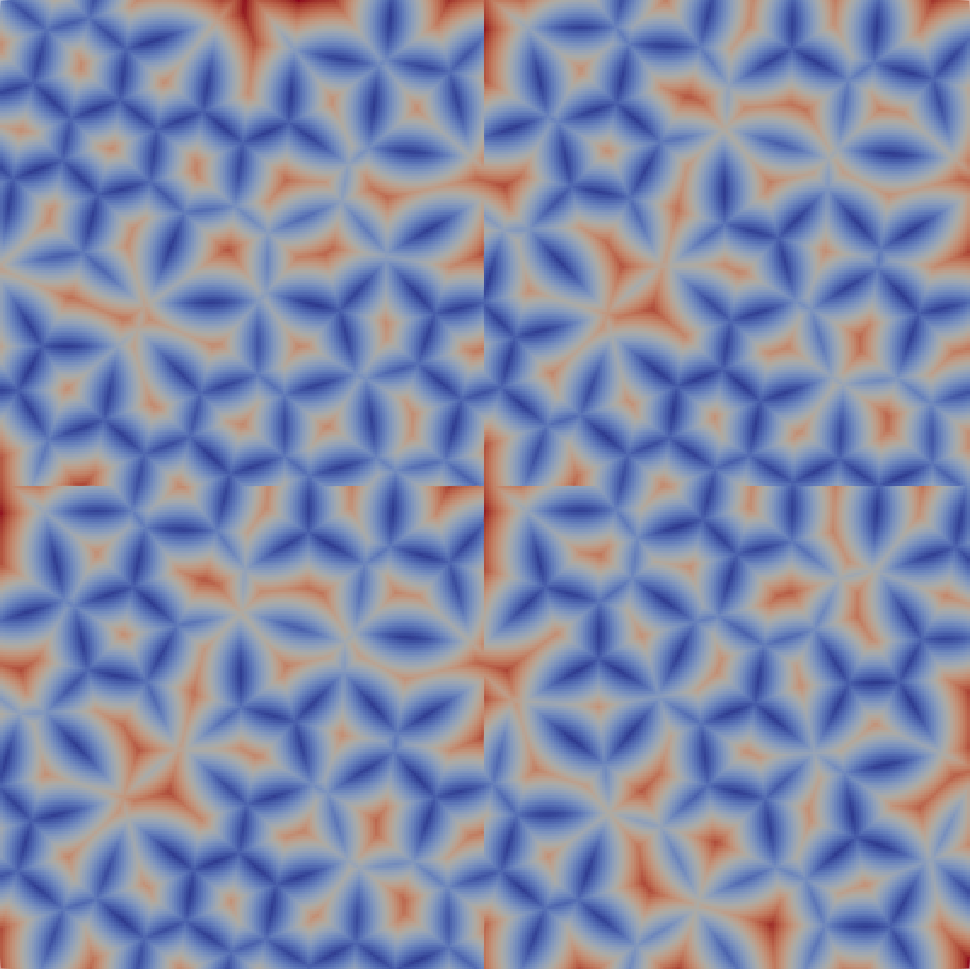} & \includegraphics[width=0.3\columnwidth]{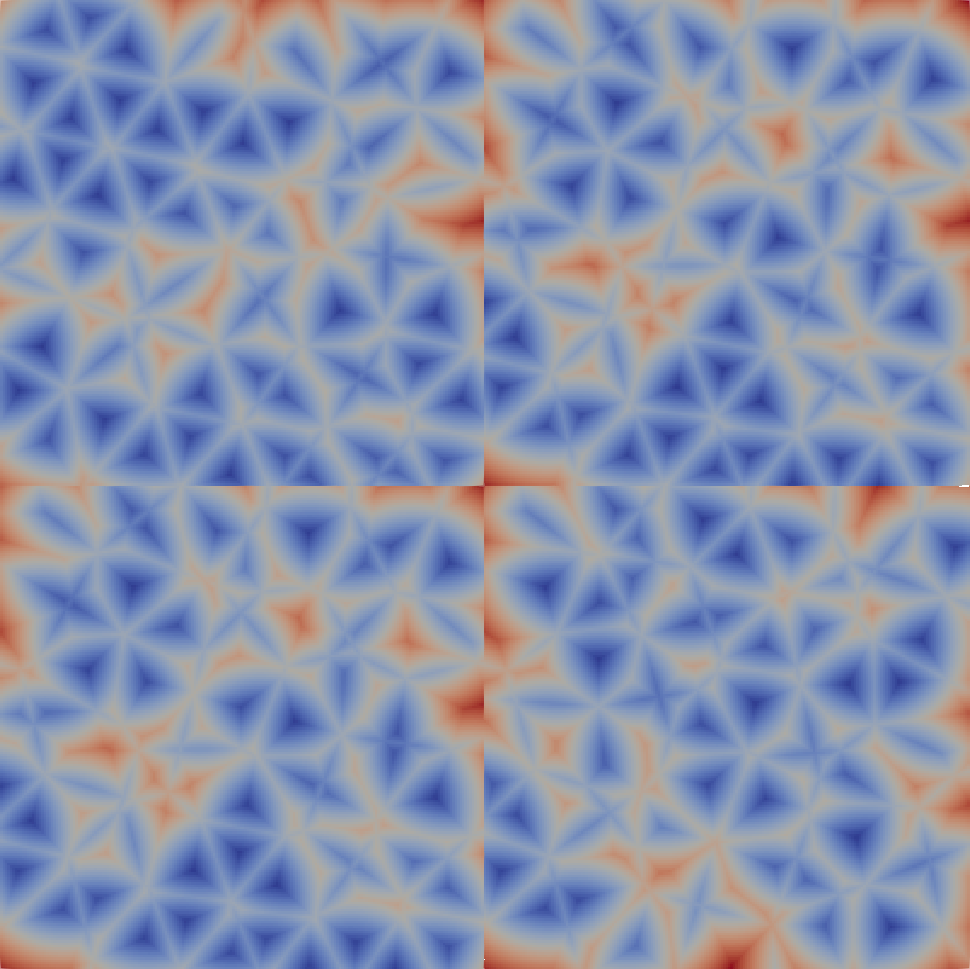}\\
		\includegraphics[width=0.3\columnwidth]{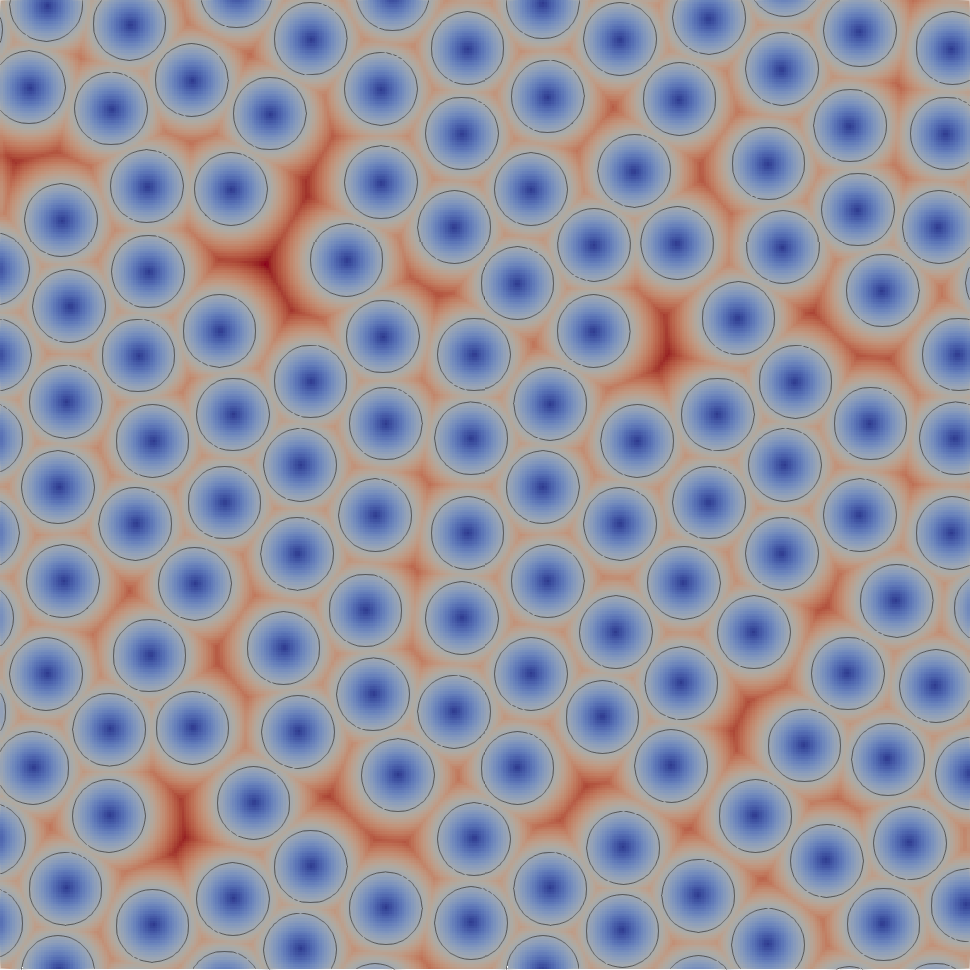} & \includegraphics[width=0.3\columnwidth]{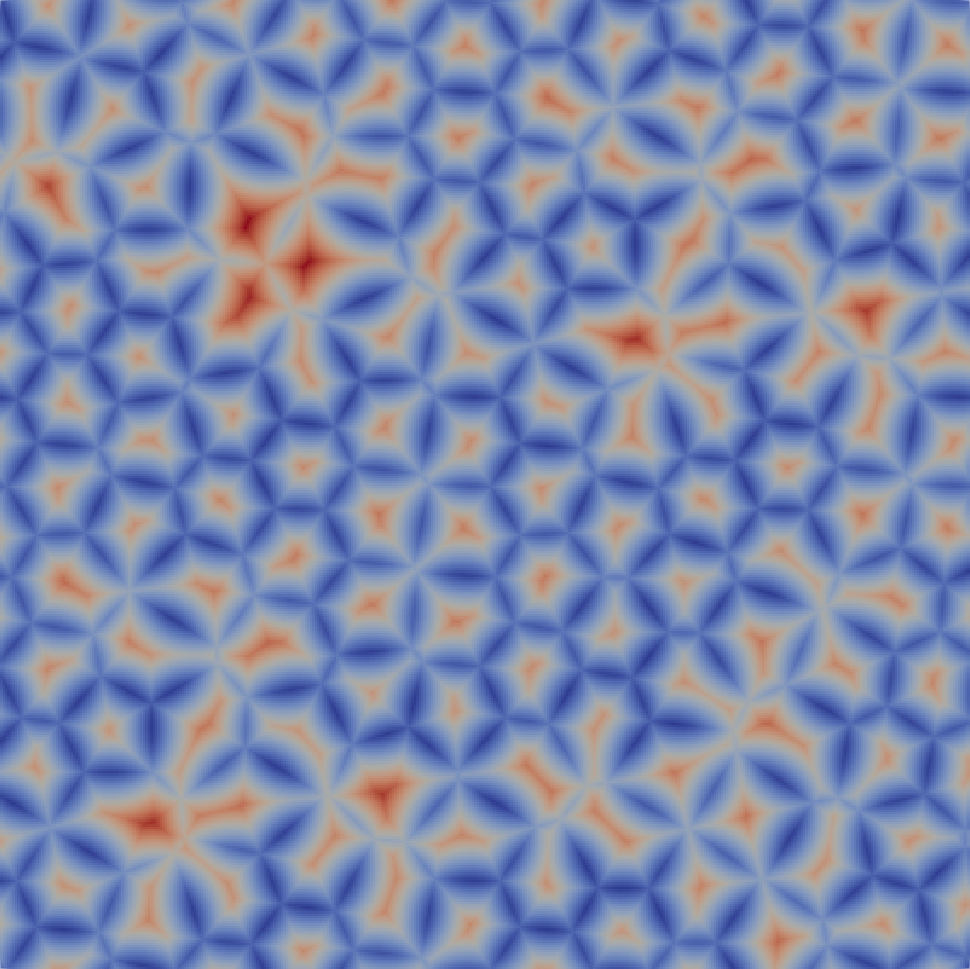} & \includegraphics[width=0.3\columnwidth]{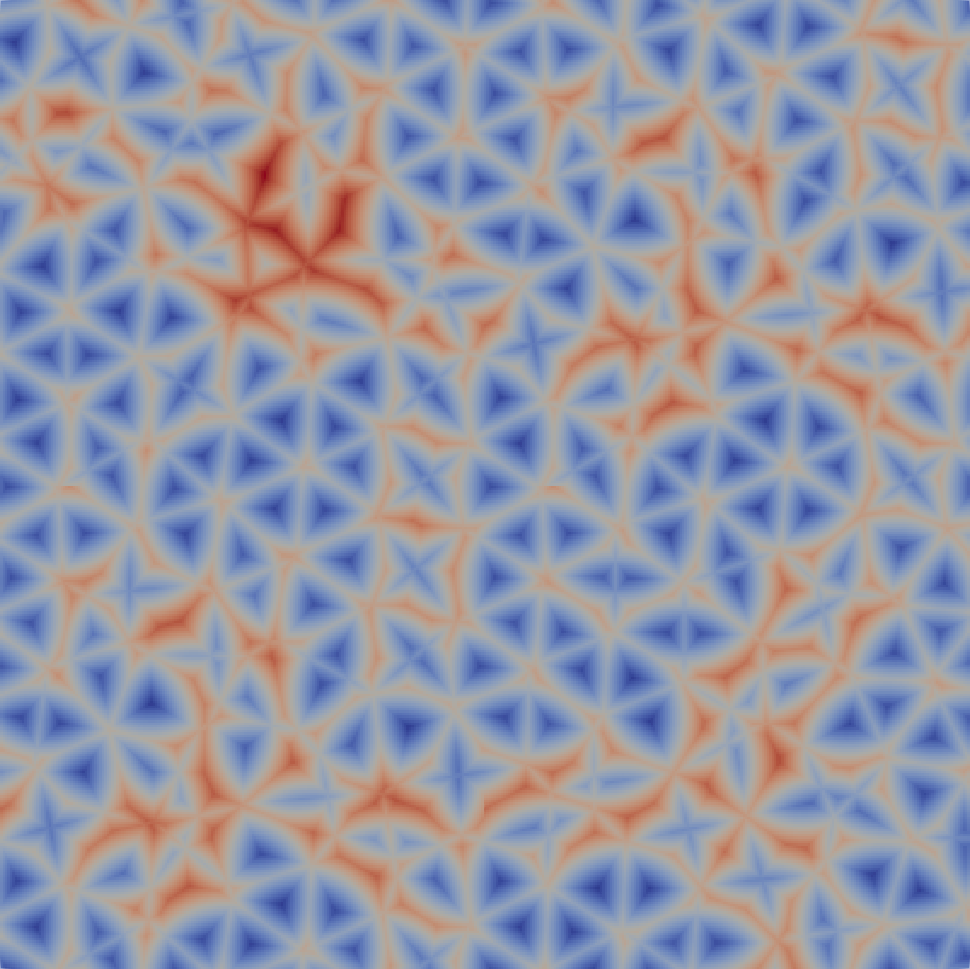}\\
		(a) & (b) & (c)
	\end{tabular}
	\caption{Assembled level-set fields composed of $2\times2$ tiles without (top row) and with (bottom row) a virtual boundary inset taken into account during particle placement. Note that even the nearest-neighbour distance $\LS$ (a) is not continuous across vertices when the virtual inset is not considered, albeit the particle boundaries, i.e. zero-value contours outlined in grey, are. Severe discontinuities then appear for the second-nearest $\LS_{\text{\romannumber{2}}}$ (b) and the third-nearest  $\LS_{\text{\romannumber{3}}}$ (c) neighbour distance without the inset boundaries. }
	\label{fig:virtualboundary}
\end{figure}

It is critical when extending the original framework to Wang tiles, with respect to the morphing procedures, to ensure that all three fields are continuous across the relevant tile edges/faces. As illustrated in~\Fref{fig:virtualboundary}, copying particle images introduced in the previous subsection is insufficient in this regard. 
To guarantee the required continuity, we define a wider domain margin as a subdomain related to individual copy-inducers. Instead of computing intersections with a tile boundary, we compute them for a virtual inset boundary (inset by $r$ is usually enough) and copy the particles according to the virtual copy-inducers. Consequently, wider portions of tile domains are restrained near boundaries, resulting in restored continuity. This modification is also reflected in the construction of auxiliary field~$\widetilde{\LS}^{\tileset}$.

\section{Results}
\label{sec:results}

\subsection{Comparison of 16-tile sets}
\label{sec:set_comparison}
\begin{figure*}[h!]
	\centering
	\setlength{\tabcolsep}{0pt}
	\begin{tabular}{cp{0.31\textwidth}cp{0.31\textwidth}cp{0.31\textwidth}}
		(a) & \includegraphics[width=0.28\textwidth]{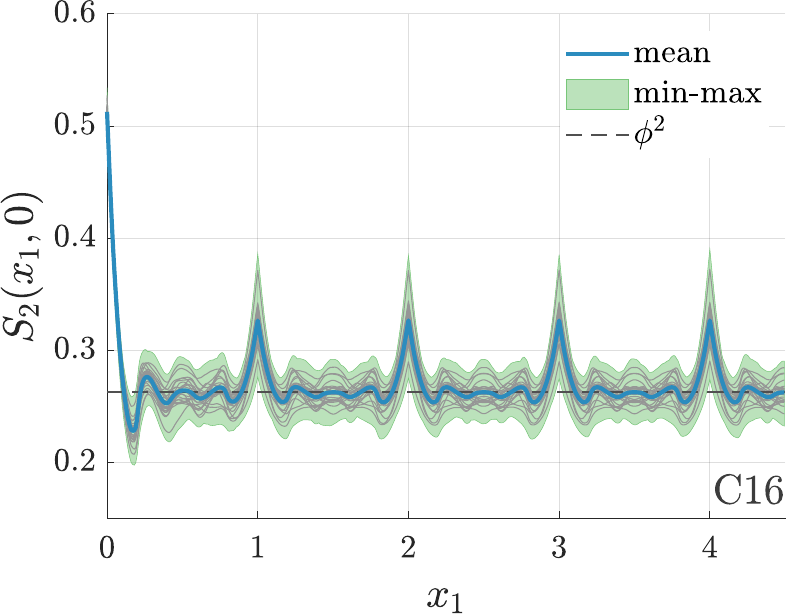} &
		(b) & \includegraphics[width=0.28\textwidth]{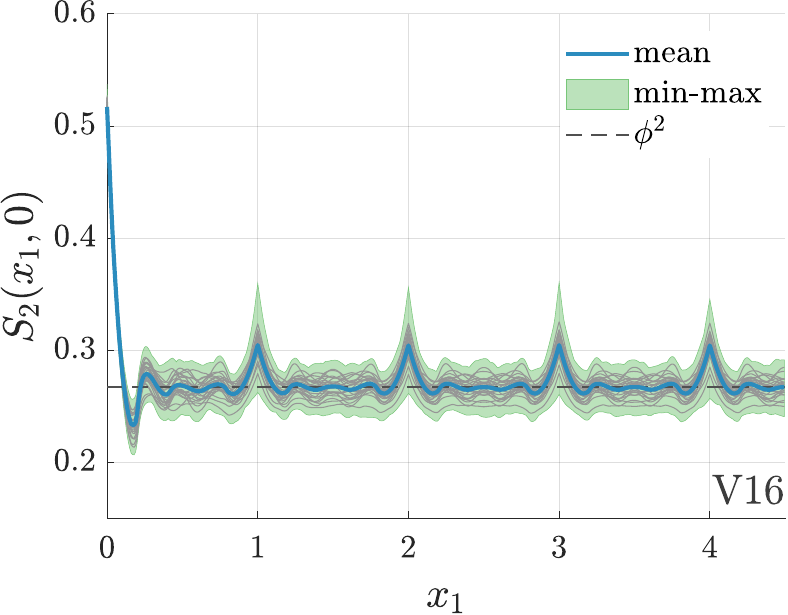} &
		(c) & \includegraphics[width=0.28\textwidth]{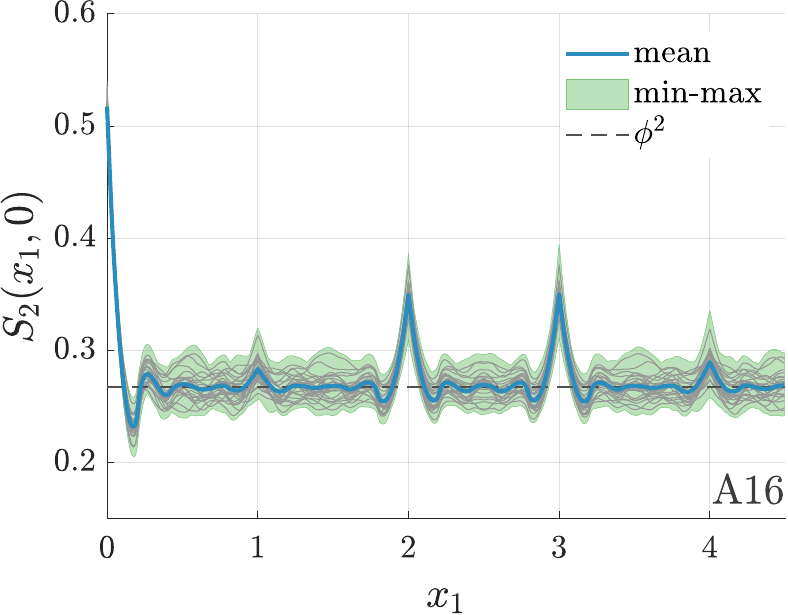}
	\end{tabular}
	\caption{$\Stwo$ cross-sections along an $x_1$ coordinate for $10\times10$ tilings for (a) the edge-defined tile set C16, (b) the vertex-defined set V16, and (c) the Amman set A16, considered in~\Sref{sec:code_analysis}. Thin grey lines depict the average over 100 realizations of each tile set morphology, solid blue line shows the overall average (i.e. over realizations and morphologies) and the green area captures the range between minimal and maximal values. Dashed grey line marks the asymptotic value $\volfrac^2$, where the average volume fraction $\volfrac$ is considered.}
	\label{fig:S2sections}
\end{figure*}
First, we compare the periodicity reduction in three planar sets depicted in~\Fref{fig:sets}, which have the same cardinality but different tile definitions. We supplement our earlier observations~\cite{doskar_aperiodic_2014} regarding suppressed artificial periodicity compared to PUC-based reconstructions and better fitness of stochastic tile sets over their aperiodic counterparts.
Namely, we consider: (C16) an edge-based stochastic set over four colours,\footnote{This set corresponds to the set reported in~\cite{cohen_wang_2003}.} (V16) a vertex-based stochastic set over eight colours, and (A16) the aperiodic Ammann's set~\cite{ammann_aperiodic_1992} over 12 colours\footnote{For the aperiodic assembly, we used the substitution rule from~\cite{stam_aperiodic_1997}.}.

Assuming ergodicity of a microstructure, we quantify the artificial periodicity in the tiling assemblies by means of the secondary peaks in the two-point probability function $\Stwo\atx$~\cite{torquato_random_2002}.
$\Stwo\atx$ states the chance of finding two points separated by $\x$ in the same phase. Hence, $\Stwo\atx$ attains its maximum at $\x = \tens{0}$, where it equals the volume fraction $\volfrac$ of a chosen phase. For microstructures without any internal ordering, $\Stwo\atx \to \volfrac^2$ with $\lVert \x \rVert \to \infty$ because two sufficiently distant points are uncorrelated. On the other hand, if a microstructure is composed of a repeating (SE)PUC, $\Stwo\atx$ exhibits secondary peaks having the same magnitude as $\Stwo(\tens{0})$ at nodes of a regular grid whose spacing corresponds to the (SE)PUC size. Thus, in the intermediate case of Wang tiling, the magnitude of secondary peaks indicates the remaining artificial ordering in the reconstructed microstructure, see~\cite{novak_compressing_2012,doskar_aperiodic_2014} for details.
For computing $\Stwo\atx$, we used the Fast Fourier Transform (FFT) instead of random sampling, because the bias in the statistics caused by the implicit periodicity introduced with FFT is negligible for reasonably large microstructural realizations~\cite{gajdosik_qualitative_2006}. Recall also that the two-point probability function $\Stwo$, widely used for statistical quantification of material microstructures, e.g.~\cite{torquato_random_2002}, is the inverse Fourier Transform image of the power spectral density (or its estimate via periodogram) traditionally used in the Computer Graphics community, e.g.,~\cite{cohen_wang_2003,kopf_recursive_2006,lagae_poisson_2006}.

For each of the three tile sets we ran the presented algorithm 25 times and thus generated 25 different tile set morphologies. Aiming at capturing the artificial periodicity related to the limited number of edge codes and tiles, we resorted to the simplest set-up with circular inclusion of a fixed radius $r=0.1$ in order to minimize the influence of particle shapes. In addition, we prevented the particles from overlapping tile vertices in order not to favour the vertex-based tile set a priori\footnote{From \Fref{fig:graphs} it follows that a particle will be copied to all tiles if it overlaps any vertex for the A16 and C16 tile sets.}. The level-set fields of each unit-size tile ($\domain = \left[-0.5,0.5\right]^2$) were discretized using a regular grid comprising $201\times201$ points. Finally, we posed two constraints, $\kappa=0.01$ and $\rho=0.03$, on $\admissibleDomain$, recall \Eref{eq:admissible_domain_contraints}.
\begin{figure}[h!]
	\centering
	\includegraphics[width=0.75\columnwidth]{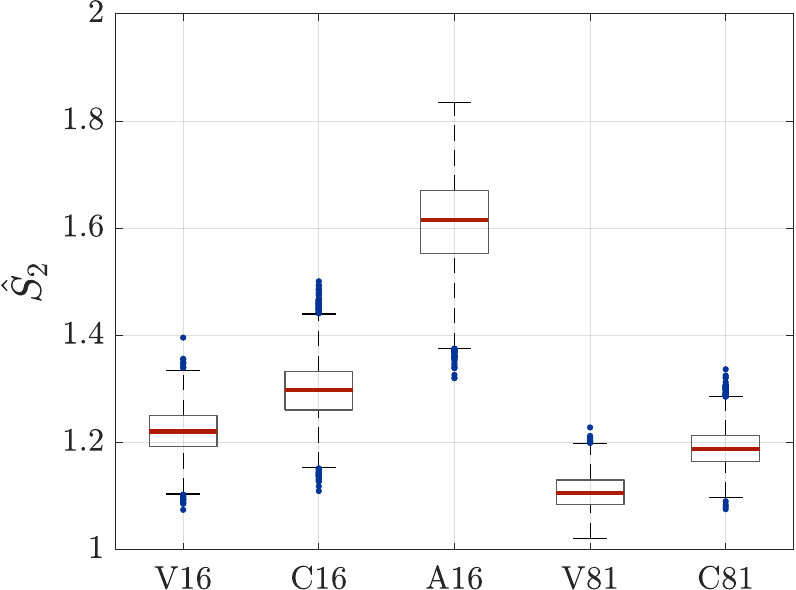}
	\caption{Box-and-whisker plot of the normalized maximal secondary peaks $\StwoSec$ over 100 realizations and 25 tile set morphologies for each tile set. The central lines mark the median; edges of each box denote the first and third quartiles; and whisker lines contain all data not considered as outliers (marked as dark blue dots).}
	\label{fig:S2peaks}
\end{figure}

We assembled 100 microstructural realizations, i.e. tilings, containing $10\times10$ tiles for each tile set and tile morphology and computed $\Stwo$ statistics for the inclusion phase. Cross-sections of $\Stwo$ along the $x_1$ coordinate are plotted in~\Fref{fig:S2sections} for the tile sets considered. As expected, the secondary peaks appear at loci with integer coordinates. Excluding the primary peak, we picked the second highest extreme $\StwoSec$ normalised against the averaged asymptotic value $\volfrac^2$ for each realisation and plotted the obtained data with box-and-whisker diagrams in~\Fref{fig:S2peaks}.

While the Ammann's tile set A16 seems promising locally---note the suppressed secondary peaks at $x_1=1.0$ and $x_1=4.0$ in~\Fref{fig:S2sections}c---its deterministic aperiodic structure leads to pronounced secondary extremes in other regions, e.g. $x_1=3.0$. This observation corroborates our previous conclusions, based solely on the distribution of individual tiles within tilings, that strictly aperodic sets such as the Ammann's or Culik's set considered in~\cite{doskar_aperiodic_2014} lead to higher secondary peaks than stochastic sets~\cite{cohen_wang_2003} with the same cardinality, which feature more uniform secondary peaks.
Figs.~\ref{fig:S2sections} and \ref{fig:S2peaks} clearly show that, out of the two stochastic sets, the vertex-based definition performs better as expected, even despite the prevented vertex overlaps. This superiority of vertex-based tile sets in two dimensions is further supported by additional results comparing vertex- and edge-based sets comprising 81 tiles, see~\Fref{fig:S2peaks}.

\subsection{2D example}
\label{sec:2Ddemo}

Second, we present an example of a two-dimensional microstructure based on polygonal particles. The geometry of each particle was derived from an originally regular, randomly-oriented polygon with the number of vertices being sampled from a normal distribution with the mean value $0.6$ and the standard deviation $0.5$, $\mathcal{N}\left(6.0,0.5^2\right)$, and rounded to the nearest integer. 
The angle between two rays connecting the particle centre and neighbouring vertices was perturbed with a value randomly chosen from $\mathcal{N}\left(0.0,0.5^2\right)$. Finally, each vertex was placed on its corresponding ray at a distance  sampled from $\mathcal{N}\left(0.95,0.05^2\right)$ (and capped by $1.0$) relative to the particle's circumscribed radius.

Following the outcomes of~\Sref{sec:set_comparison}, we picked the vertex-based set with 16 tiles depicted in~\Fref{fig:sets}b. Similarly to the previous set-up, the size of each tile was $\domain = \left[-0.5,0.5\right]^2$ and the corresponding level-set fields were discretized using a regular grid comprising $201\times201$ points.
The initial value of the circumscribed radius was set to $\radius = 0.08$. After 40 algorithm steps, the radius was increased to $0.1$, and eventually we reduced it to 0.06 after next 60 steps. The number of algorithm steps and related radii were chosen randomly to illustrate the algorithm's control options.
Only the first two constraints on $\admissibleDomain$ from \Eref{eq:admissible_domain_contraints} were posed. We fixed the minimal particle distance at $\kappa=0.01$; the maximal distance $\rho$ was set to $0.05$ in the first 50 steps and decreased to $0.02$ later. The width of the virtual boundary inset was kept constant at $0.1$.
\begin{figure*}[ht!]
	\centering
	\begin{tabular}{cc}
		\includegraphics[height=6cm]{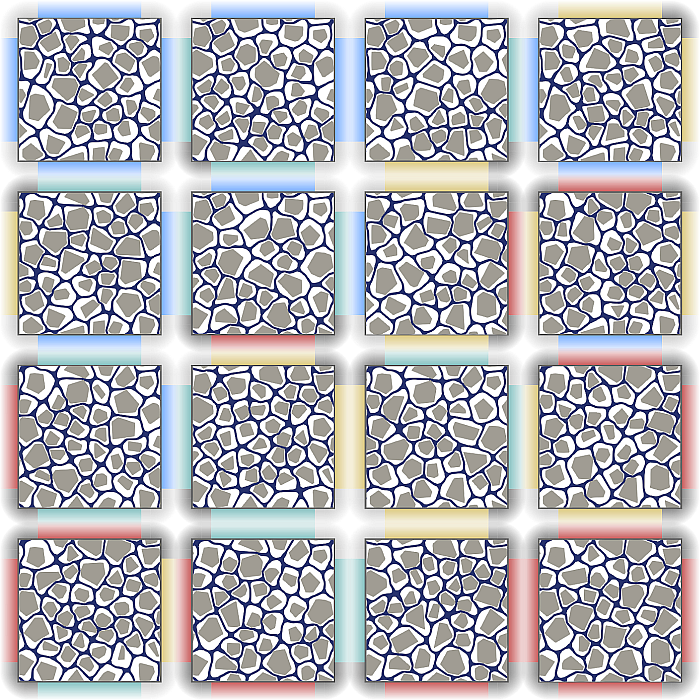} & \includegraphics[height=6cm]{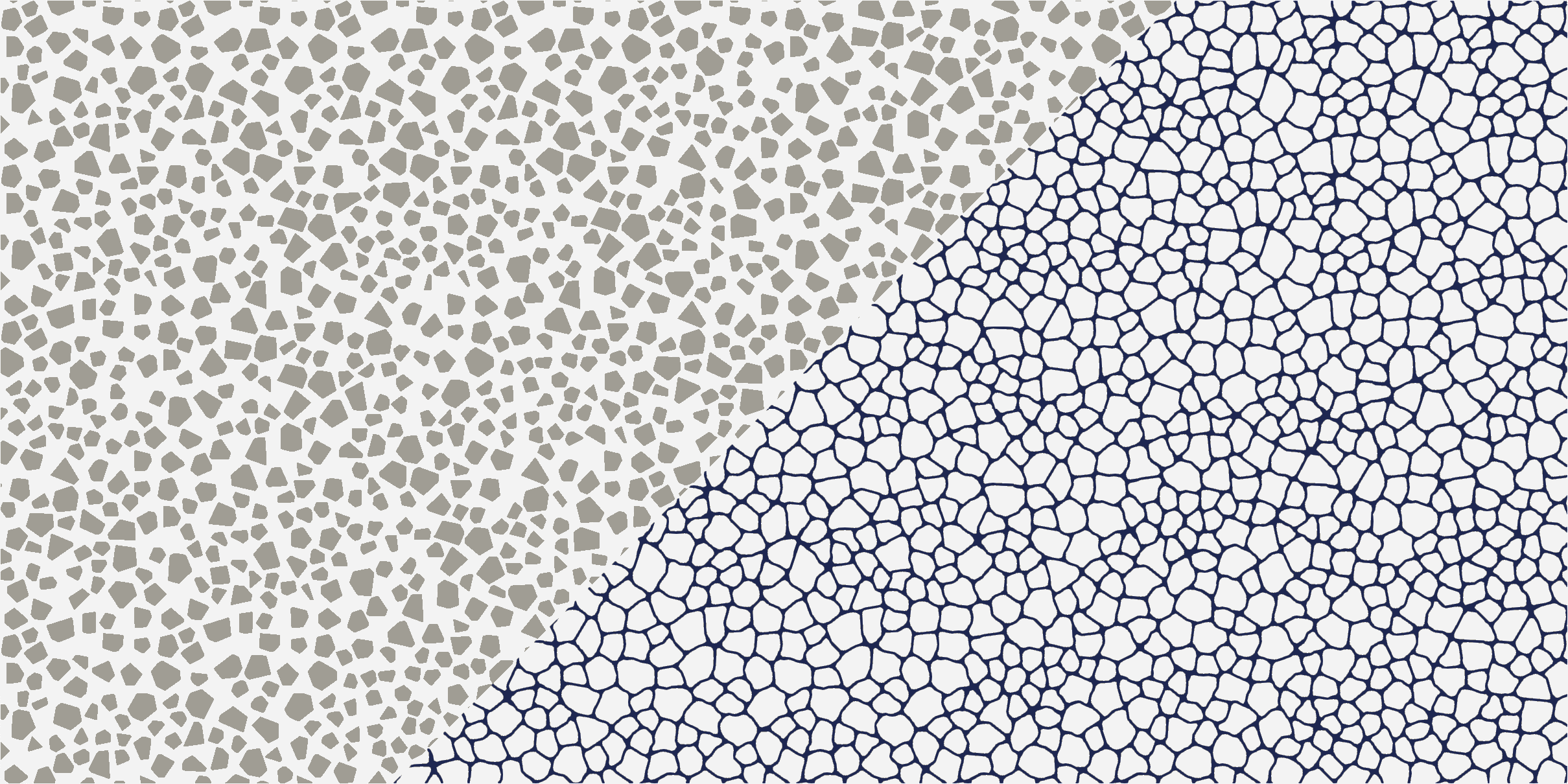} \\		 
		(a) & (b)
	\end{tabular}
	\caption{Results of the algorithm set-up described in~\Sref{sec:2Ddemo}: (a) A composed view of the tile set with highlighted edge and vertex codes. The particulate assembly is drawn in grey; the corresponding foam-like microstructure is shown in blue (similar to \Fref{fig:morphing}); (b) An example of a $10 \times 5$ tiling.}
	\label{fig:example2D}
\end{figure*}

\Fref{fig:example2D} depicts a composed view of the resulting particle assembly and the related closed-cell, foam-like microstructural geometry obtained by considering $t_{\textrm{c}} = 0.015$ and $t_{\textrm{o}} = 0.020$ in~\Eref{eq:foam_composition}.

\subsection{3D example}
\label{sec:3Ddemo}

Finally, we demonstrate a three-dimensional output of the algorithm for a set of 16 Wang cubes with two codes for each face orientation uniformly distributed in the set. 

Again, we used centred tile domains of a unit size discretized with a grid of $101\times101\times101$ points.
For the packing part of the algorithm, we considered ellipsoidal particles with a random orientation and a fixed circumscribed radius $r=0.10$. The ratio between the middle and the major semi-axes lengths was sampled from a uniform distribution $\mathcal{U}\left(0.7,0.9\right)$; the minor to the major semi-axes length ratio was randomly generated from $\mathcal{U}\left(0.6,0.7\right)$. The admissible domain at each algorithm step was dictated both by $\LSs{}$ and $\LSs{2}$ with constraints $\kappa = 0.02$, $\rho = 0.05$, and $\sigma=0.05$ in~\Eref{eq:admissible_domain_contraints}. Since we aimed at modelling a foam-like microstructure, the width of a virtual inset boundaries was set to $0.1$.

\begin{figure}
	\centering
	\begin{tabular}{c}
		\includegraphics[width=0.95\columnwidth]{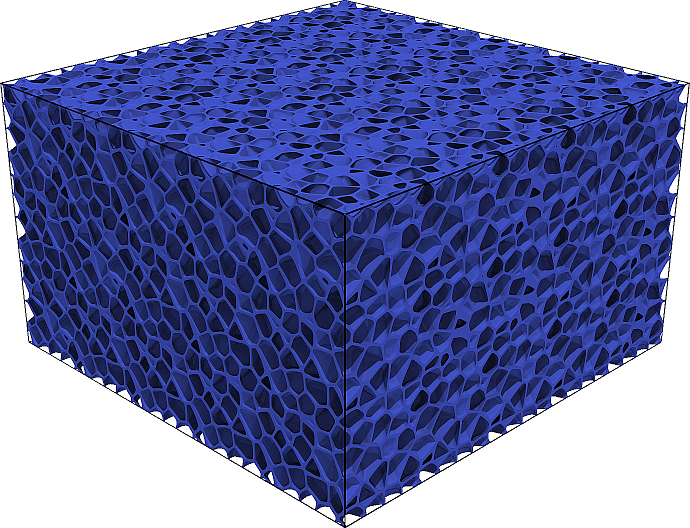}
	\end{tabular}
	\caption{A microstructural realization assembled as a $5\times5\times3$ tiling from the Wang tiles generated in~\Sref{sec:3Ddemo}. }
	\label{fig:example3D}
\end{figure}

The final microstructure was obtained by performing the morphing operations according to~\Eref{eq:foam_composition} with $t_{\textrm{c}} = 0.02$ and $t_{\textrm{o}} = 0.03$.
For the sake of brevity, we do not show individual Wang cubes, and we plot only a microstructural sample comprising $5\times5\times3$ tiles in~\Fref{fig:example3D}.

\subsection{Computational cost}
\label{sec:timing}

\begin{figure}[h!]
	\setlength{\tabcolsep}{-0pt}
	\begin{tabular}{cc}
	(a) & \includegraphics[width=0.85\columnwidth]{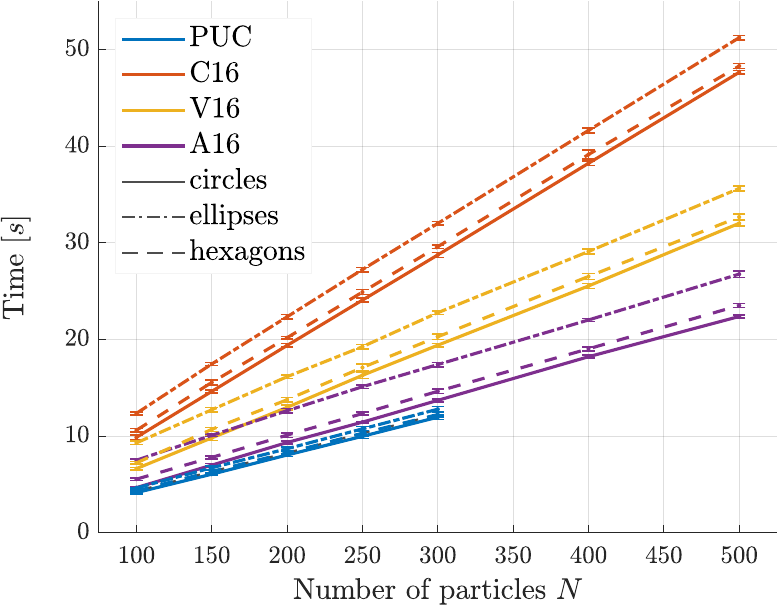}\\
	(b) & \includegraphics[width=0.85\columnwidth]{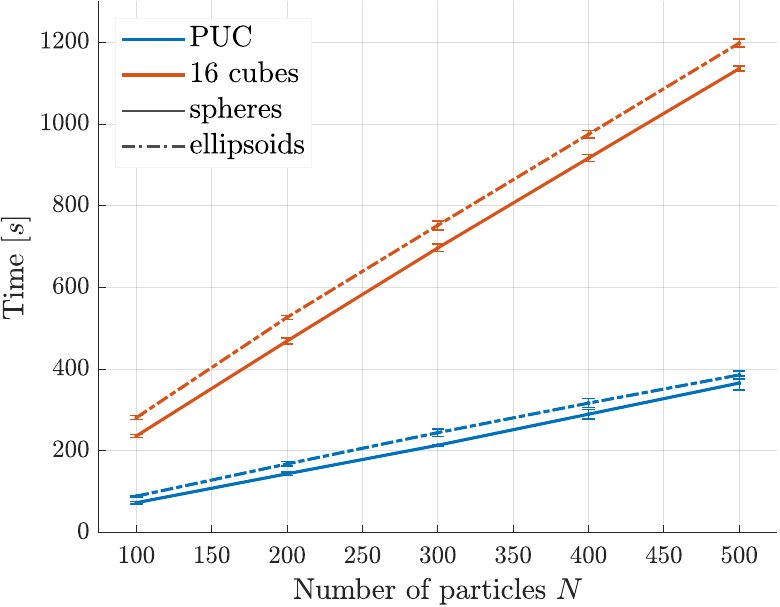}
	\end{tabular}
	\caption{Wall-clock time needed to pack $N$ particles into (a) two-dimensional and (b) three-dimensional tile sets. Results for different particle shapes are drawn in distinct line styles; considered tile sets (including the trivial case of PUC) are distinguished with distinct colours. Error bars show the standard deviation for each setting. PUC results in two dimensions are limited by the number of particles that can be packed in a unit domain.}
	\label{fig:timing}
\end{figure}

Our extension inherits the $\mathcal{O}(N)$ complexity of the Sonon et al.'s method~\cite{sonon_unified_2012,sonon_advanced_2015}, as shown in~\Fref{fig:timing}. However, the actual time needed to generate particle packings or complex morphologies depends on several factors. While the impact of the grid resolution is straightforward (computational time scales linearly with the number of grid points), the effect of other factors such as particle shape and compatibility constraints due to different definitions of a tile set is less obvious.

To illustrate the influence of these factors, wall-clock time for several settings is plotted in \Fref{fig:timing}. Tile sets considered for the analysis are the two-dimensional ones used in \Sref{sec:set_comparison} and the three-dimensional sets from \Sref{sec:3Ddemo}; in addition, a PUC was included in the comparison as a trivial case of a tile set. For all settings, tile domains of unit size were discretized with $401\times401$ grid points in two and $101\times101\times101$ grid points in three dimensions. The circumscribed radius of particles was set to $\radius = 0.02$ for two-dimensional and $\radius = 0.05$ for three-dimensional problems. The algorithm was terminated once the desired number of placed particles was achieved. Computational time\footnote{All computational times are reported for a workstation equipped with an Intel\textsuperscript{\textregistered} Xeon\textsuperscript{\textregistered} E31280 3.50 GHz processor and 16 GB RAM running Windows 10 version 1903. The algorithm was implemented in the latest standard of C++ language (C++17), relying on parallel algorithms from the C++'s Standard Template Library.} for each setting was measured $10$ times.

Interestingly, albeit of the same cardinality, individual tile sets from~{\Sref{sec:set_comparison}} exhibit significantly different computational times. This is due to the distribution of edge codes within a set, because a larger part of the tile set domain must be updated with each particle copy in sets with fewer codes, recall~{\Sref{sec:extension_tiles}}. Note that Ammann's set with six codes for both horizontal and vertical edges behaves nearly the same as PUC.

Computational time related to the morphing operations depends on the operational mode of the algorithm. If the algorithm is used for both packing and morphing at the same time, the level-set fields used during the particle packing phase can be directly utilized for the morphing operations. On the other hand, when the morphing part runs separately, all the distance fields must be calculated again. Yet this recalculation is typically faster than the packing because all copies and compatibility constraints are already taken into account; for instance, generating the particle packing for the microstructure presented in \Sref{sec:3Ddemo} took 1,280~s while the recalculation needed for the morphing operations finished in 420~s.

\section{Summary}

In this work, we have extended Sonon et al.'s level-set based framework~\citep{sonon_unified_2012} to the microstructural models based on the formalism of Wang tiles. The extension enables generating compressed representations of complex microstructural geometries such as open and closed foams or cells, which have been nearly impossible~\cite{doskar_aperiodic_2014} or very expensive~\cite{novak_compressing_2012} to generate for Wang tiles to date.

Advancing from the standard Periodic Unit Cell to the generalized periodicity of Wang tiles necessitated several modifications of the original algorithm that addressed difficulties originally not encountered in the case of PUC.
Driven by the geometrically-motivated question of where a vertex-overlapping particle should be copied to, we have come up with a simple procedure based on a graph analysis capable of revealing the underlying vertex definition of a tile set, if present. 
We have also demonstrated that a straightforward copying of boundary-intersecting particles according to the edge and vertex codes is insufficient for preventing spurious particle overlaps because the information about the interior of a tile is not automatically communicated across individual tiles. As a remedy, we have introduced an artificially updated level-set field that facilitates this necessary communication.
The width of the updated region in this field is dictated by the required final geometry. If only a particle distribution is to be generated, the width equal to the particle radius $\radius$ suffices. On the other hand, if foam-like microstructures are desired, the virtual boundary inset by $\radius$ is necessary in order to preserve continuity of the level-set fields $\LSs{2}$ and $\LSs{3}$ across tile edges, resulting in the updated region of width $2\radius$.

Additionally, we have devised two minor modifications, one capable of breaking a regular underlying grid of possible particle centres by introducing their random shifts within an admissible region, the other accelerating the $\LSs{}$ updates by pre-computing $\LSp$ on a patch. The importance of the latter modification grows with higher $\LSs{}$ resolutions and larger cardinality of a tile set.

Having a universal framework for generating Wang tile morphologies at our disposal, we have supplemented our previous comparison of strictly aperiodic and stochastic tile sets~\cite{doskar_aperiodic_2014} and confirmed the superiority of vertex-defined tile sets in suppressing artificial periodicity in assembled microstructural samples. Given the same cardinality of the edge- and vertex-based tile sets in two dimensions, we recommend using the latter. On the other hand, the same comparison in three dimensions is more subtle and the choice should be always a compromise between available computational resources and required periodicity reduction, following e.g. the approximate formula proposed in~\cite{novak_compressing_2012}.

Admittedly, the presented modification inherits certain limitations posed by the original framework; namely, it provides only indirect control of a resulting microstructural geometry and no spatial statistics are involved in the particle placing process. Albeit addressing these limitations remained out of the scope of this work, ideas conceptually similar to \cite{yang_new_2018}, i.e. optimizing the particle positions a posteriori to minimize the discrepancy between target and computed spatial statistics, could be potentially applied to steer the generated microstructure to the desired statistics.

Albeit our extension features linear complexity as in the original Sonon et al.'s framework, the generalized periodicity brings increased computational cost compared to PUC generation, see~\Sref{sec:timing}. Recall though that tile-based compression is intended primarily for generating numerous stochastic realizations with arbitrary sizes. Thus, more time can be spent on preparing the initial microstructure because the increased cost will be amortized later during repeated use of the same compression.

Our current focus is on developing a robust finite element discretization tool that would enable meshing both outputs of the framework, i.e. analytical geometries of particulate microstructures and complex geometries implicitly defined via level-set fields and processed with the marching-cube algorithm. In both cases, special attention will be paid to ensuring topological and geometrical compatibility of the resulting finite-element meshes across the corresponding faces/edges.

\section*{Acknowledgment}
M. Doškář, J. Novák, and D. Rypl acknowledge the endowment of the Ministry of Industry and Trade of the Czech Republic, project No.~FV10202. 
This research was performed at the Center of Advanced Applied Sciences (CAAS), financially supported by the European Regional Development Fund (project No.~CZ.02.1.01/0.0/0.0/16\_019/0000778).
M. Doškář was also partially supported by the Grant Agency of the Czech Technical University in Prague through student project No. SGS18/036/OHK1/1T/11, and J. Zeman received partial support from Czech Science Foundation project No. 17-04301S.
We thank Stephanie Krueger for helpful comments and proof-reading the manuscript.

\bibliography{bibliography}
\bibliographystyle{my-elsarticle-num-names}
\biboptions{square,numbers,sort&compress}

\end{document}